\documentclass[10pt,compsoc]{IEEEtran}
\ifCLASSOPTIONcompsoc
  \usepackage[nocompress]{cite}
\else
  \usepackage{cite}
\fi
\usepackage{enumitem}
\usepackage{placeins}

\ifCLASSINFOpdf
\else
\fi

\usepackage{float}
\usepackage{url}
\usepackage{amsmath}
\usepackage{subcaption}

\usepackage{graphicx}
\usepackage[usenames,dvipsnames]{xcolor}
\newcommand{\nbh}[1]{\textsf{#1}}

\usepackage{placeins}

\usepackage[numbers]{natbib}
\usepackage{subcaption}
\usepackage{booktabs}
 
\usepackage{fancyhdr}

\begin{document}
\pagestyle{fancy}
\fancyfoot[CE, LO]{\vspace{8px}\small Preprint of Herremans D., Chew E.. 2017. MorpheuS: generating structured music with constrained patterns and tension. \emph{IEEE Transactions on Affective Computing}. PP(99). DOI: 10.1109/TAFFC.2017.2737984}

\title{MorpheuS: generating structured music with constrained patterns and tension}

\author{Dorien Herremans,~\IEEEmembership{Senior Member,~IEEE,}
        and Elaine Chew,~\IEEEmembership{Member,~IEEE,}
 \IEEEcompsocitemizethanks{ \IEEEcompsocthanksitem E. Chew is with the Centre for Digital Music, School of Electronic Engineering and Computer Science, Queen Mary University of London,  UK.\protect\\
\IEEEcompsocthanksitem 
D. Herremans was at the Centre for Digital music (see above) for most of the research, and is currently with the Information Systems Technology and Design Pillar at Singapore University of Technology and Design.\protect\\
Email: see http://dorienherremans.com/contact}
\thanks{Manuscript received 000, 2016; revised 000, 20xx.}}

%
%

\markboth{IEEE Transactions on Affective Computing}%
{MorpheuS: generating structured music with constrained patterns and tension}
%



\IEEEtitleabstractindextext{%
\begin{abstract}
Automatic music generation systems have gained in popularity and sophistication as advances in cloud computing have enabled large-scale complex computations such as deep models and optimization algorithms on personal devices. Yet, they still face an important challenge, that of long-term structure, which is key to conveying a sense of musical coherence. We present the MorpheuS music generation system designed to tackle this problem. MorpheuS' novel framework has the ability to generate polyphonic pieces with a given tension profile and long- and short-term repeated pattern structures. A mathematical model for tonal tension quantifies the tension profile and state-of-the-art pattern detection algorithms extract repeated patterns in a template piece. An efficient optimization metaheuristic, variable neighborhood search, generates music by assigning pitches that best fit the prescribed tension profile to the template rhythm while hard constraining long-term structure through the detected patterns. This ability to generate affective music with specific tension profile and long-term structure is particularly useful in a game or film music context. Music generated by the MorpheuS system has been performed live in concerts.
\end{abstract}


\begin{IEEEkeywords}
Affective Computing, Music, Music retrieval and generation, Affective computing applications, Sound and Music Computing, Entertainment, Variable Neighborhood Search, Pattern Recognition

\end{IEEEkeywords}}



\maketitle

\IEEEdisplaynontitleabstractindextext

\ifCLASSOPTIONpeerreview
\begin{center} \bfseries EDICS Category: 1-ASLAS, 2-OPTI, 1-CPRS, 9-MSAM, 9-STCM \end{center}
 \fi
%
\IEEEpeerreviewmaketitle

\IEEEraisesectionheading{\section{Introduction}\label{sec:introduction}}



%
%
%
%

\IEEEPARstart{T}{echnologies} for digital music have become increasingly important, bolstered by rising global expenditures in digital music in excess of 64 billion USD in 2014 alone~\cite{mckinsey2015}. The popularity and relevance of automatic \emph{music generation} has recently been underscored by the launch of Google's Magenta project\footnote{\url{https://magenta.tensorflow.org/welcome-to-magenta}}, ``a research project to advance the state of the art in machine intelligence for music and art generation''.  In this research, we develop a music generation system, called Morpheus~\cite{herremans2016morpheus}, that tackles one of the biggest remaining challenges in the field of automatic music composition: long term structure. Long term structure is that which generates coherence over larger time scales from phrases up to the entire piece; it refers to more than simply traditional ABA form, and includes the modulation of features such as loudness and tension, and the use of recurrent patterns and motivic transformations over time, so as to generate coherence over these large time scales. While most existing music generation systems create pieces that may sound good over a short time span, these outputs often lack long-term structure. 
MorpheuS can take any existing polyphonic music piece as template and morph it into a new piece with a predefined tension profile. The new piece will also preserve the same long term structure (i.e. pattern structure) as the template. 

To this day, it remains notoriously difficult to enforce constraints (e.g. long-term structure) in music generation systems based on machine learning methods such as Markov models~\cite{pachet2011markov}. 
In previous research, the first author therefore developed a novel method for constraining long-term structure through an optimization-based approach, combined with machine learning. The proposed framework consisted of an efficient variable neighborhood search (VNS) optimization algorithm that is able to generate melodies (or monophonic music) with a fixed semiotic structure (e.g. AABBACA)~\cite{herremans2013composing, herremans2014thesis, herremans2015generating} and evaluates its solution through the Euclidean distance between a Markov model built on a corpus and one trained on the generated piece. This research shows that the approach offers a viable way of constraining structure. In the current paper, the VNS algorithm is expanded to generate \emph{complex polyphonic music}. Although the algorithm is able to work with any type of polyphonic music, as a proof of concept, we focus on piano music in this research. 


A second novel aspect of MorpheuS is the inclusion of an original \emph{tension} model. Tension shapes our music listening experience. In some music genres, such as game and film music, there often is a direct relationship between tension, a narrative, and the emotion perceived~\cite{cohen2001music}. In this research, we aim to generate music that has a predefined tension profile through time. The target tension can be specified by the user, or calculated from a template piece using a computational model developed by the authors~\cite{Herremans_tenor2016}. A system like Morpheus that can generate music having a particular tension profile could be employed by film makers, composers, and game programmers when music matching a specific narrative is desired. 

The third contribution of this research is the integration of a state-of-the-art pattern detection algorithm~\cite{meredith2013cosiatec}, which is used to find recurring patterns and themes in the template piece. MorpheuS then uses the patterns found to configure the structure in a newly generated piece by introducing the patterns as hard-constraints during the generation process. 

\begin{figure}[ht]
\centering
 \includegraphics[clip, trim=2cm 3.8cm 8.4cm 1.24cm, width=0.48\textwidth]{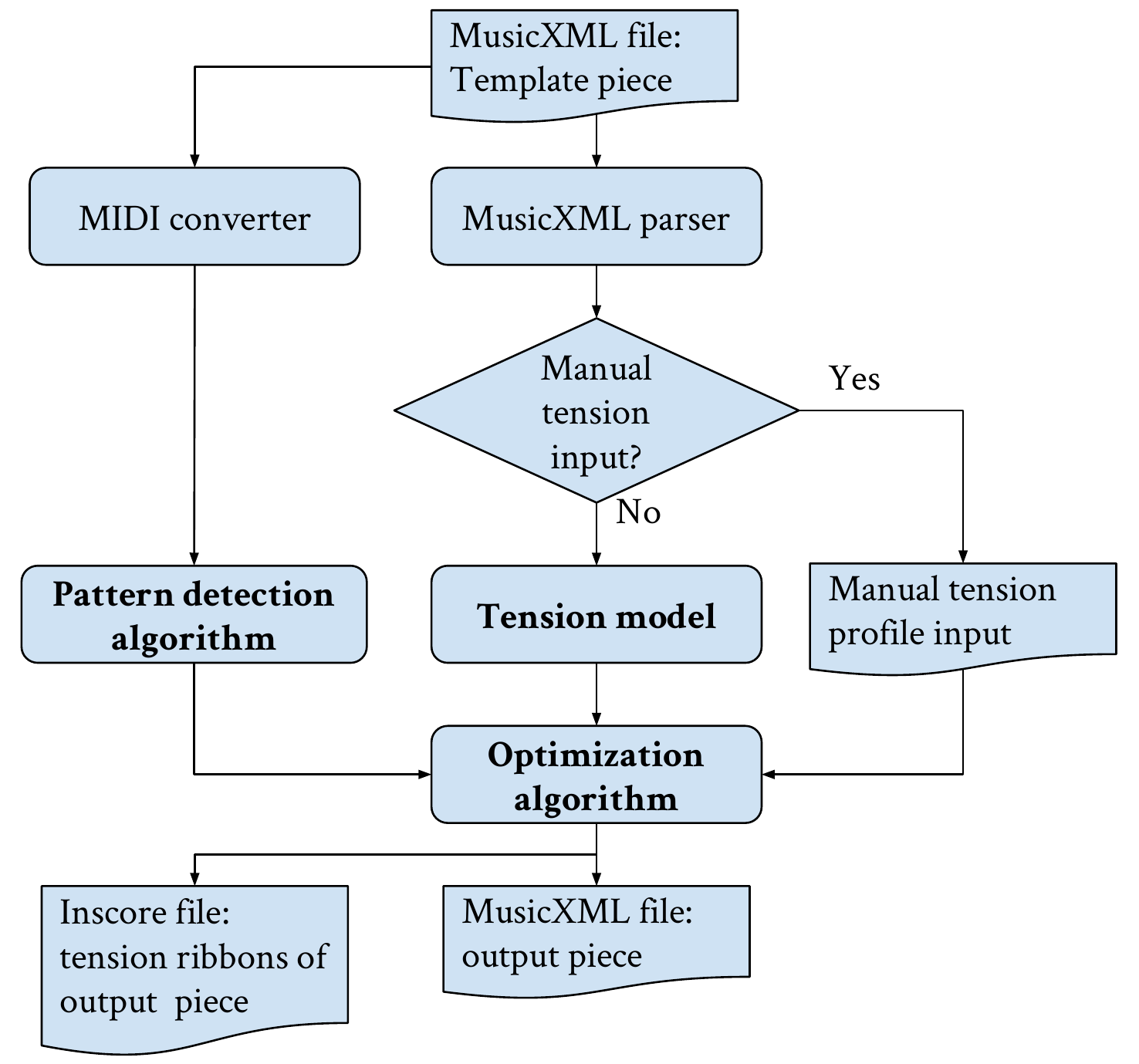}
 \caption{Overview of MorhpeuS' architecture  \cite{herremans2016morpheus}.}
 \label{fig:arch}
\end{figure}

MorpheuS' functional architecture is displayed in Figure~\ref{fig:arch}. The system is implemented entirely in Java. The main modules of the algorithm, in bold, will be further discussed in Sections~\ref{sec:tensionmodel},~\ref{sec:patterndetection}, and~\ref{sec:opt}. Before embarking on these discussions, we briefly survey related systems in the next section.

\section{Literature review}


Before examining the individual components of the MorpheuS system, we give an overview of related research. The first subsection covers different techniques used in automatic music generation; the focus of this overview lies mainly on metaheuristic optimization algorithms. 
Next, we focus on the two types of long term structure that are incorporated in the system. 
A first aspect of long-term structure is ensured through a tension profile. By requiring that the music adhere to a coherent tension profile, MorpheuS can generate music displaying specific tension 
characteristics throughout the piece. This makes the output particularly well suited to game or film music scenarios. An overview thus follows of previous research on music generation with a narrative and tension. 
We then address the second aspect of long-term structure in MorpheuS, namely, recurring patterns in generated music, providing a review of how this has previously been tackled by researchers in automatic composition. For a more general survey of current music generation systems, the reader is referred to~\cite{herremans2017taxonomy, fernandez2013ai}.



\subsection{Generation techniques for music}

The idea that computers could compose music is as old as the computer itself. Ada Lovelace, who worked with Charles Babbage on the Difference Engine, predicted that the engine when realised could one day ``compose elaborate and scientific pieces of music of any degree of complexity or extent''~\cite{lovelace1843notes}.

Since then, many automatic systems for music generation have been developed. In the 50s, the first piece composed entirely by a computer, ``The Illiac Suite'', was generated by a stochastic rule-based system~\cite{hiller1957musical}. More recently, a number of systems based on Markov models were developed for simple melody generation~\cite{pinkerton1956information, conklin1995multiple}, to harmonization~\cite{pachet2001musical, chuan2011generating} and improvisation systems \cite{dubnov2012music, assayag2006omax, franccois2013mimi4x}. In recent years deep learning models have entered the scene \cite{eck2002first, chen2001creating, ICML2012BoulangerLewandowski_590, cancino2017bach, sabathe2017deep, huang2016chordripple, hutchings2017drums, herremans2017modeling}. While many of these systems produce output that sounds good on a note-to-note level, they often lack long-term coherence. We aim to tackle this challenge in this research by employing pattern detection techniques. In order to exploit the patterns found, we opt for an optimization-based approach, which allows us to constrain structure. 

In Section~\ref{sec:opt} the problem of generating music with long-term structure is defined as a combinatorial optimization problem. This problem is computationally complex to solve, as the number of possible solutions grows exponentially with the length of the piece. As an example, a piece consisting of only 32 notes, with 24 possible pitches per note, has $32^{24}$ possible solutions. 

There have only been limited attempts at solving music generation problems with \emph{exact methods} such as integer programming. For example, \citet{cunha2016} uses integer programming with structural constraints to generate guitar solos based on existing licks. Their objective function is based on music theoretic rules. In research by~\cite{tanaka2015describing}, the authors propose a method to generate counterpoint---independent linear voices that combine to form a single harmonic texture. They formulate the generation task as an integer programming problem that uses existing composition rules as constraints to control global structure.  However, this work remains a theoretical formulation, with no solution method as yet implemented.


In order to overcome the difficulty and often long computational run times required to calculate exact solutions to optimization problems, many practical applications use \emph{metaheuristics}. A metaheuristic is defined by~\citet{sorensen2015metaheuristics} as ``a high-level problem-independent algorithmic framework that provides a set of guidelines or strategies to develop heuristic optimization algorithms. The term is also used to refer to a problem-specific implementation of a heuristic optimization algorithm according to the guidelines expressed in such a framework.'' These techniques often employ a variety of strategies to find a good solution in a limited amount of computing time; they do not guarantee an optimal solution, but typically good solutions are found~\cite{blum2003metaheuristics}.

There exist three main groups of metaheuristics: population-based, constructive, and search-based algorithms~\cite{sorensenmetaheuristics}. The first group, which includes evolutionary algorithms, has seen recent gains in popularity in the literature. Population-based algorithms get their name from the fact that they inter-combine a set of solutions (population) to create new ones.~\citet{horner1991genetic} were the first to develop a genetic algorithm for music generation. These techniques have later been used to generate jazz solos \cite{biles2001autonomous}, counterpoint style music \cite{mcintyre1994bach, polito1997musica, phon1999evolving}, and rhythmic patterns \cite{tokui2000music, horowitz1994generating}, and to combine fragments for orchestration \cite{carpentier2010solving}. 

The second group, constructive metaheuristics, gradually build solutions from their constituent parts, for example, by growing individual notes in sequence. An example of this category includes ant colony optimization, which was first applied to music in 2007 to harmonize baroque music~\cite{geis2007ant}. 

The third category, local search-based heuristics, typically make iterative improvements to a single solution. They include algorithms such as iterated local search, simulated annealing, and variable neighborhood search~\cite{sorensenmetaheuristics}. An example of these techniques in the field of music generation can be found in the research of~\citet{davismoon2010combining}, who used simulated annealing to generate music according to a fitness function that was derived from a Markov model. The first author of the current paper, was the first to develop a variable neighborhood search (VNS) algorithm that was able to generate counterpoint music~\cite{herremans2012composing, herremans2014thesis}. This VNS was shown to outperform a genetic algorithm on the same task and has been modified in the current research to generate complex polyphonic music. 


\subsection{Narrative and tension}


The tension profile, which is integrated in the algorithm so as to shape the tension of the generated music, is particularly important when generating music with a narrative, or program music. Program music has a long and illustrious history, a well-known example being Richard Strauss' ``Don Quixote''. Such narrative music tells a story, by using a set of organizational, representational, and discursive cues that deliver story information to the audience. Such cues can include tension profiles, leitmotifs (recurring melodic fragments associated with a person, idea, or story situation), and others. All of these elements typically elicit varying emotion responses during the unfolding of a piece when synchronized with simultaneous media such as video or game play. Existing systems in the domain of video and game music are discussed, followed by a more focused overview of literature on tension models. 

\subsubsection{Generating film music}

A prominent application of music with narrative is film music. Music has been shown to be an important source of perceived emotion in film~\cite{cohen2001music, parke2007quantitative}. 
While~\citet{prechtl2014methodological} has conducted research on generating music that evokes basic emotions in the context of games, very little research exists on developing music generation systems that follow the emotion content of films. Even commercial applications such as the web-based music generation app, Jukedeck\footnote{\url{jukedeck.com}}, do not yet take into account the emotion narrative. Jukedeck generates background music for YouTube videos using a combination of rules and deep learning. 

A prototype system that generates background music and sound effects for short animation films was developed by~\citet{nakamura1994automatic}. For each scene, music (harmony, melody, rhythm) is generated based on rules from music theory whilst taking into consideration the mood, the intensity of the mood, and the musical key of the previous scene. The sound effects are determined by the characteristics and intensity of the movements on screen. In the next subsection we will discuss the related topic of game music.

\subsubsection{Game music -- blending}

The most dynamic form of narrative in music can be found in computer games, whereby a user creates his or her own unique scenario when moving through the game. The accompanying music needs to follow and support the suspense and emotion of the current game play.
Game music is rarely generated on the fly. Short audio files are generally cross-faded together as the player moves through different game states~\citep{collins2008game}. An exception to this common practice can be seen in the game, Depression Quest\footnote{\url{https://isaacschankler.bandcamp.com/album/depression-quest-ost}}, as the music is generated dynamically as the user moves through the different scenarios of the game. 
With current cross-fading techniques, it is not uncommon for two fragments to clash rhythmically or harmonically, causing a jarring change in the music. The automatic DJ-system developed by~\citet{muller2012data} ensures smooth blending, yet the audio fragments need to be harmonically and rhythmically similar for the blending to work successfully. By restricting the range of harmonies and rhythms in this way, one also limits the musical variations and expressive capacity of the music. To overcome this limitation, some games implement procedural music approaches that use control logics or rules that control playback. One of the first procedural music and audio approaches for computer games can be found in the game `Otocky' for the Famicom platform. Otocky is a side-scrolling shooter, whereby the player controls a ship that fires balls at both enemies and flying musical notes. The melody part is formed by the player's firing behavior and in-game mechanics, and is rendered on top of a two note bass line~\cite{collins2009introduction}. For an overview of procedural music techniques, the reader is referred to \citet{collins2009introduction}. 

More recent work has focused on incorporating elements of tension and emotion into adaptive game music. \citet{prechtl2016adaptive} created a system that generates music from scratch instead of using existing fragments. Prechtl uses a Markov model for chord generation that takes into account emotion parameters such as alarm or danger. His study uncovered correlation between mode and valence, and between tempo/velocity and arousal.

\citet{casella2001magenta} created MAgentA (not to be confused with Google's music generation project Magenta), an abstract framework for a video game background music generation that aims to create ``film-like'' music based on the mood of the environment using a cognitive model. At the time of publication, the authors mention that the framework was being integrated from the abstract level into the FantasyA Virtual Environment, but no further references could be found.

The system developed by~\citet{brown2012mezzo} makes use of the concept of ``Leitmotifs''
, commonly used in Western opera.~\citet{brown2012mezzo}'s system stores different forms of each motif corresponding to varying degrees of harmonic tension and formal regularity. This allows the system to choose the appropriate fragment corresponding to the current state and story of a game. The reader is referred to~\citet{collins2009introduction} for a more complete overview of dynamic music in games.




\subsubsection{Generating music with tension}

Musical tension is an important tool for evoking emotion. According to \citet{farbood2012parametric}, the way that different listening experiences translate into actual `affect' is a complex process. Musical tension, measured based on musical structures, provides a stepping stone to understanding and quantifying subjective, emotional responses. The link between emotion and tension has become apparent in many studies~\cite{sloboda1991music, krumhansl1997can, scherer2001emotional, steinbeis2006role}. 
If a music generation system can generate music according to given tension profiles, it becomes directly relevant for applications in game and film music. Recent research has made advances in the quantification of aspects of musical tension, such as tonal tension~\cite{lerdahl2007modeling, Herremans_tenor2016}, even combining them to produce a composite model~\cite{farbood2006quantitative}. Based on an extensive empirical experiment,~\citet{farbood2012parametric} built a tension model that takes into account multiple musical parameters to obtain one comprehensive tension rating. Farbood implemented an earlier version of her tension model, that does not yet integrate features,~\citep{farbood2006quantitative} in the graphical computer-assisted composition system called Hyperscore, in which users can intuitively edit and visualize musical structures as they compose music~\cite{farbood2007composing}. Hyperscore shows low-level and high-level musical features (such as color, shape, dynamics, harmonic tension) and maps them to graphical elements which can be controlled by the user when crafting compositions. Thus, a user can draw a tension profile and Hyperscore will generate music with a similar profile.

Similarly,~\citet{browne2009global}'s system arranges pre-written motifs according to a pre-specified tension profile using simulated annealing. An artificial neural network was used to compute tension profiles. The objective function of the algorithm was then formed by taking the Kullback-Leibler divergence between the desired and observed tension profiles. The optimal arrangement was then taken to be the one that minimizes this distance.

In this study, we will focus on multiple aspects of \emph{tonal} tension independently versus considering a composite tension characteristic. The tonal component has proven to be a particularly strong structural influence on emotions.  In~\citet{rutherford2002experiment}'s scary music study, they conclude that more scary music is generated by breaking the Western tonal music rules. This result was empirically verified by users who rated the scariness of the generated music. 
The computational model used in this research for calculating tonal tension is discussed in more detail in Section~\ref{sec:tensionmodel}. The next section  considers the importance of patterns in music.

\subsection{Structural patterns in generated music}

Music is more than just a succession of notes that only needs to sound good in the short term. Having long-term structure: motives, patterns and variations of those patterns are essential for an enjoyable and satisfying listening experience. Music generation systems that use traditional statistical sampling methods based on Markov models typically only ensure short term relationships between notes~\citep{herremans2015generating}. 

One approach to obtaining long term structure was implemented by~\citet{roig2014automatic}, whose system concatenates rhythmic and melodic patterns in order to form new melodies based on a combination of rules and a statistical method. 
More complex statistical learning methods, such as recursive neural networks have recently gained popularity in the field of music generation due to the availability of large amounts of digital music data and increased computing power. While the first neural network for melody generation was implemented in the late 80s~\cite{todd1989connectionist}, this approach has become more relevant due to the ability of these networks to learn complex relationships between notes given a large enough corpus. 
Recent research in audio transcription by~\citet{ICML2012BoulangerLewandowski_590} shows promising results for music generation as well. They use a piano roll representation for polyphonic pieces to build a Recurrent Temporal Restrictive Bolzmann Machine (RT-RBM)-based model. This model learns harmony and melody, and local temporal coherence. Long term structure is not yet captured by the model. The polyphonic music generation system designed by~\citet{lattner2016imposing} implements convolutional restricted boltzmann machines and constrains the self-similarity matrix of a generated piece to a template. 

Another approach to long-term structure is explored by \citet{herremans2015generating} who examined the integration of Markov models in an optimization algorithm. By looking at different ways a statistical model can be used to construct an objective function, the approach ensures that the generated music has the same statistical distribution of features as a target dataset of pieces. By treating the problem of music generation as an optimization problem, ~\citeauthor{herremans2015generating} were able to impose larger-scale structure (e.g. ABBAC) on the generated music, in addition to short term statistical constraints. The resulting optimization problem was solved by a VNS metaheuristic to generate music for bagana, an Ethiopian lyre. This approach is extended in the current research to polyphonic music, with automatic detection of more complex long term patterns in the template piece. The detection method is described in greater detail in Section~\ref{sec:patterndetection}, after the next section, which focuses on the tension model.

\section{Quantifying tension in music}
\label{sec:tension}
\label{sec:tensionmodel}

Tension is a composite characteristic, which makes it very hard to capture or measure in a quantitative way. According to Mary Farbood~\cite{farbood2012parametric}, ``increasing tension is a feeling of rising intensity or impending climax, while decreasing tension can be described as a feeling of relaxation or resolution'' (p. 387). In~\citet{Herremans_tenor2016}, the authors developed a model for tonal tension based on the spiral array~\cite{chew2014tonality}, a three-dimensional model for tonality. The relevant part of the model is briefly described below, together with how it was implemented in the MorpheuS system to quantify tension in an optimization context.

\subsection{The Spiral Array}

\begin{figure*}[h]
\centering
\begin{subfigure}[b]{.3\textwidth}
	\includegraphics[width=1\textwidth]{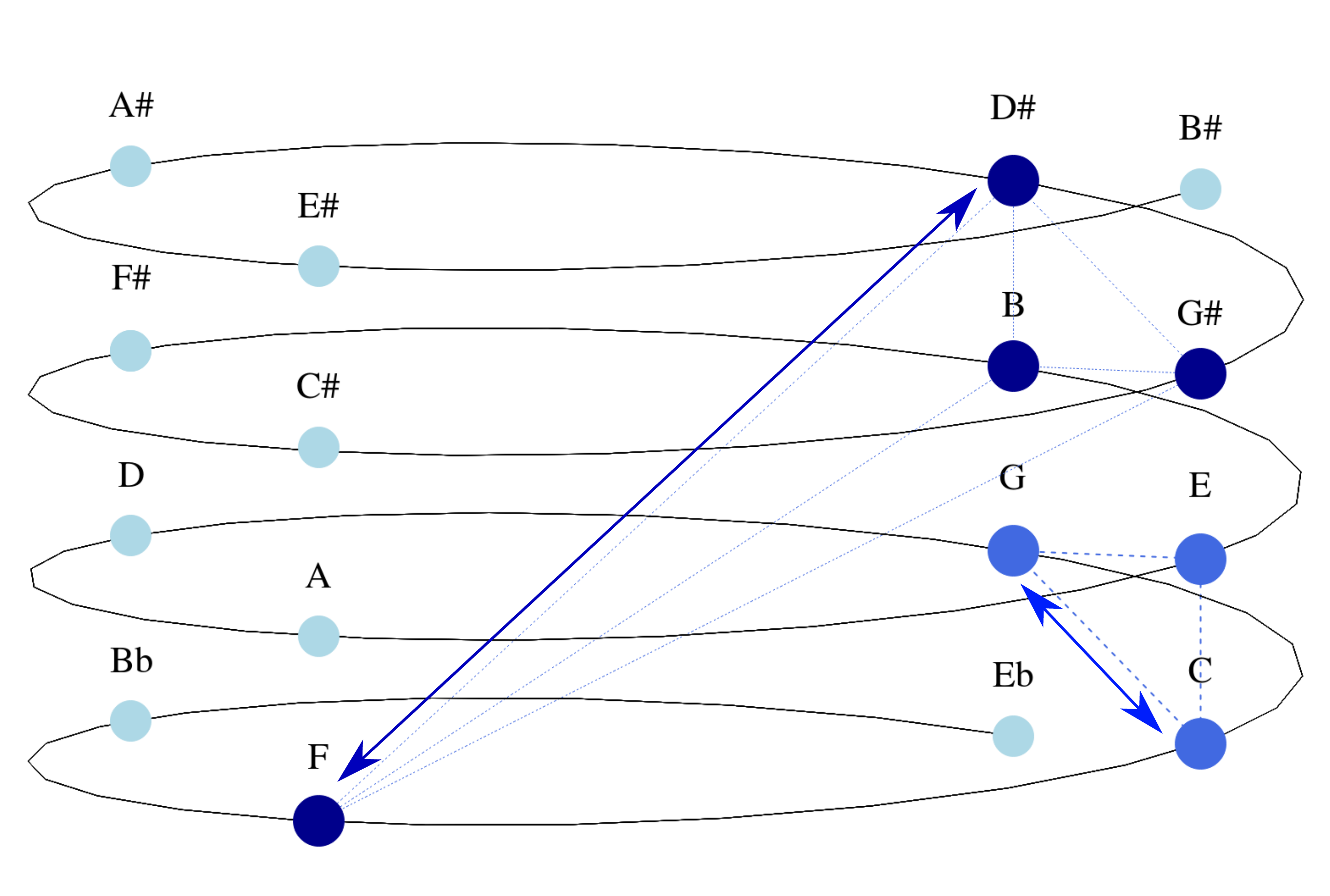}
\caption{Cloud diameter of a C major chord (small) versus the Tristan chord (large).}
\label{fig:diam}
\end{subfigure}
\begin{subfigure}[b]{.3\textwidth}
\includegraphics[width=1\textwidth]{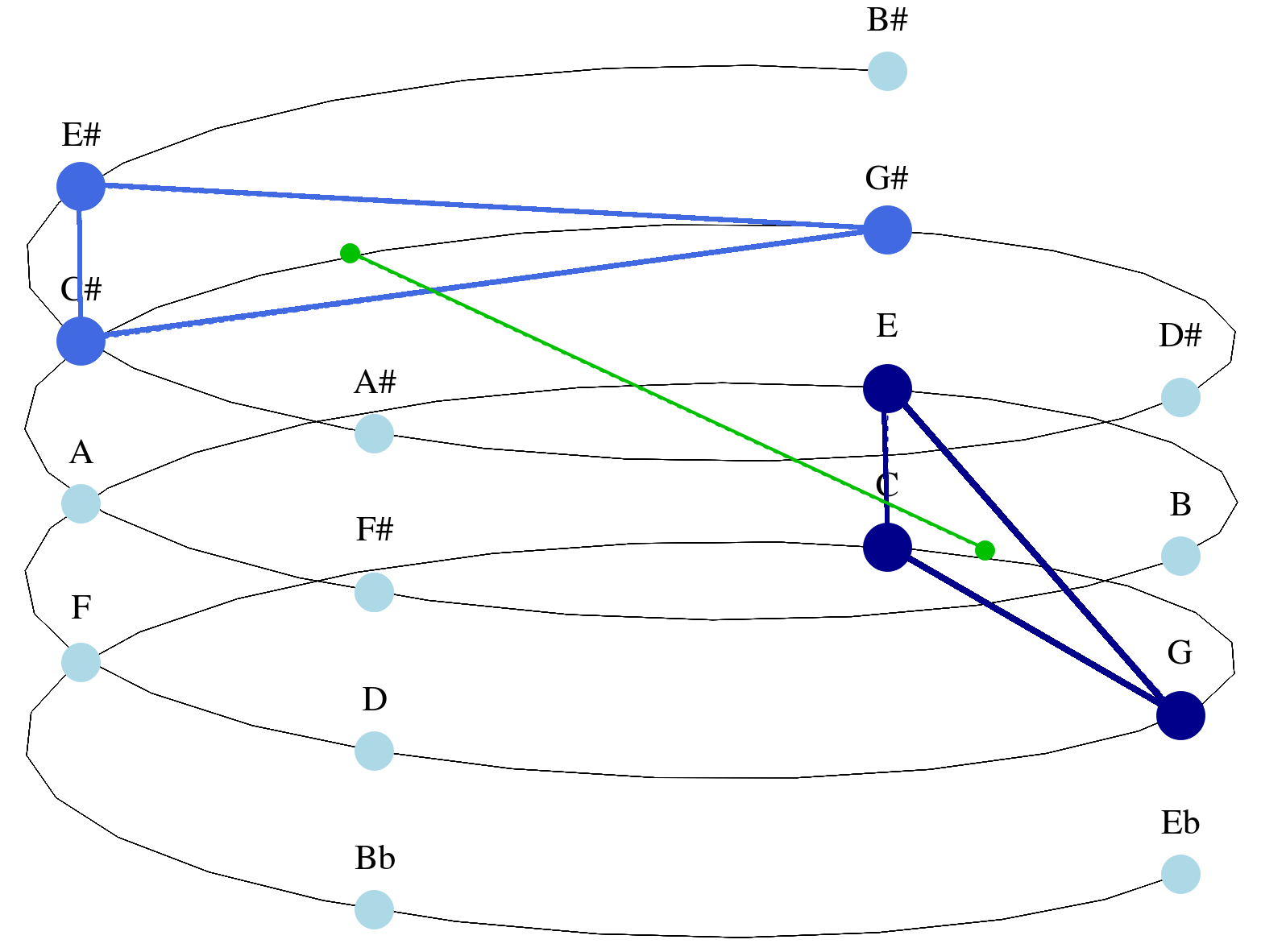}
\caption{Cloud momentum from a C major chord to a C\# major chord.}
\label{fig:mom}
\end{subfigure}
\begin{subfigure}[b]{.3\textwidth}
\includegraphics[width=1\textwidth]{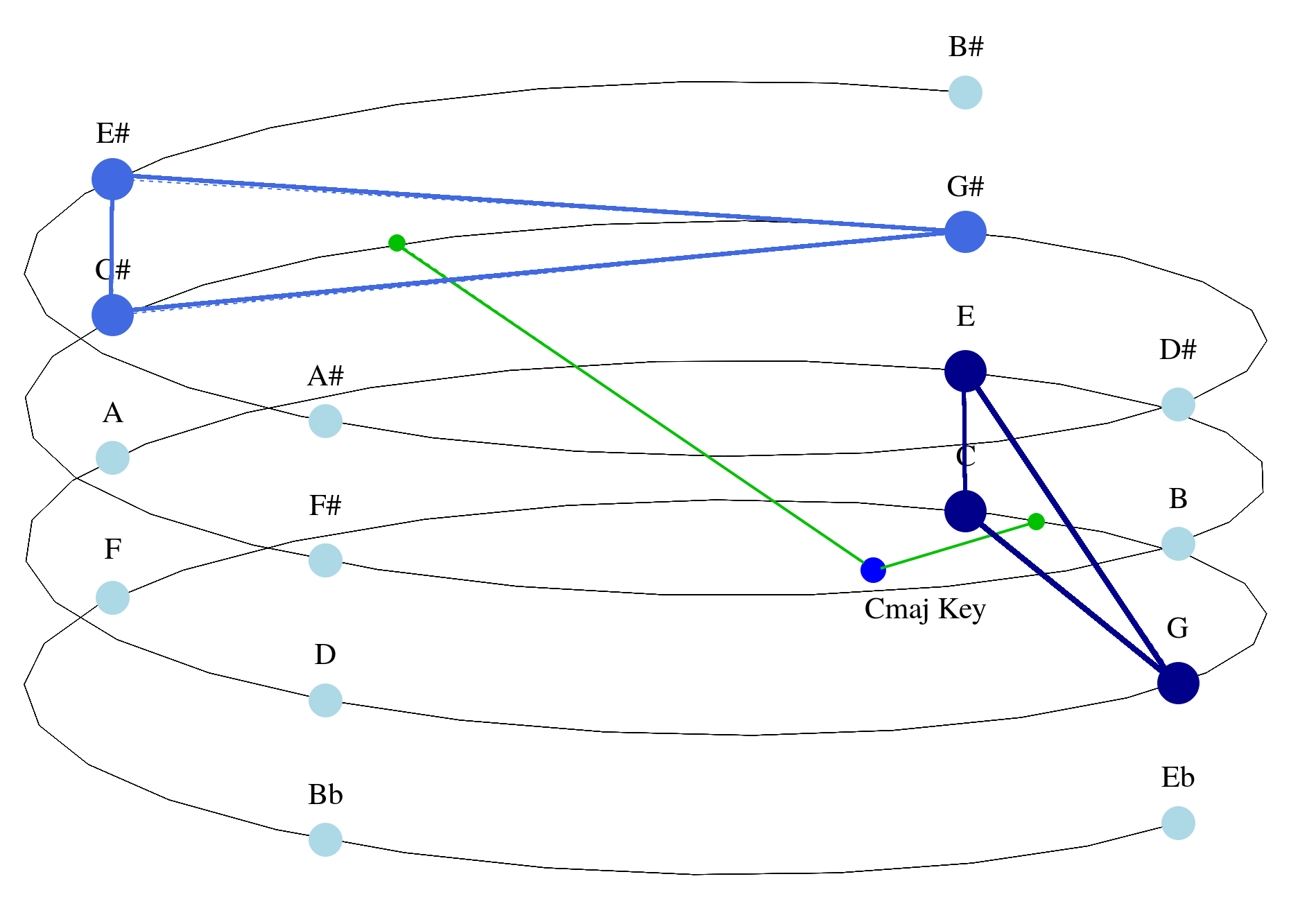}
\caption{Tensile strain of a C major and C\# major chord in the key of C major.}
\label{fig:key}
\end{subfigure}
\caption{An illustration of the three tension measures in the pitch class helix of the spiral array.}
\label{fig:spiral}
\end{figure*}

The spiral array is a three-dimensional geometric model for tonality~\cite{chew2014tonality}. It consists of an outermost helix representing pitch classes (shown in Figure~\ref{fig:spiral}), and inner helices (not shown) representing higher level tonal constructs such as chords (major and minor) and keys (major and minor) successively generated from their lower level components. Any set of points in the spiral array can be weighted and summed to generate a \textit{center of effect} (\textit{c.e.}), representing the aggregate tonal effect of its components.

Tonal representations in the spiral array mirror close tonal relationships between the entities, such as a perfect fifth between pitches, by their proximity in 3D space. For example, pitches one fifth apart are adjacent to each other in the pitch class helix (e.g. C-G) and pitches one major third apart are arranged vertically above one another (e.g. C-E). Similarly, the most likely key or chord of a cluster of pitches can be identified through a search for the key representation nearest to the c.e. of the pitch cluster. The tension model in MorpheuS uses only the pitch class and the major and minor key helices. The spiral array takes pitch spelling into account, meaning that enharmonically equivalent (but differently-spelled) pitches, such as G\# and Ab have different spatial representations. The interested reader can refer to~\cite{chew2014tonality} for a full description of the spiral array model.

\subsection{A quantitative model of tension}

The model, developed by the authors in~\cite{Herremans_tenor2016}, represents tonal tension as captured by the geometry in the spiral array. The software that calculates the tension according to this model is freely available online\footnote{\url{http://dorienherremans.com/tension}}. In order to calculate the tonal tension of a musical fragment, the piece is first divided into equal length segments, which can be mapped to clouds of points in the spiral array. The segment length is expressed in beats and can be set by the user (default setting is an $\frac{1}{8}$ note), a more detailed discussion of the effect of the segment length can be found in~\cite{Herremans_tenor2016}. Based on these clouds, three measures of tonal tension can be computed: 

\begin{description}
\item[Cloud diameter] captures the diameter of the cloud of notes, which measures the dispersion of the cloud in tonal space.  

\item[Cloud momentum] reflects the movement in tonal space between two consecutive clouds of notes, by quantifying the distance between their c.e.'s.

\item[Tensile strain] measures the distance between the c.e. of a cloud and the position of the global key in the array.

\end{description}

Figure~\ref{fig:spiral} illustrates each of the three tension measures with the pitch class helix of the spiral array. On the left, the (small) cloud diameter of a C major triad is shown together with the (much larger) diameter of the tristan chord, a well known tense chord~\cite{magee2002tristan}. The large tonal distance traversed by a transition from the C major to the C\# major chord is illustrated in Figure~\ref{fig:mom}, an example of the cloud momentum measure. Finally, Figure~\ref{fig:key} visualizes the tonal distance between the c.e.'s of each these two chords and the key of C major, which shows two contrasting tensile strain measures.

For exact mathematical details of how to calculate the three measures of tension, the reader is referred to~\cite{Herremans_tenor2016}. MorpheuS uses these three tension characteristics to evaluate the musical output and match it to given template tension profiles.  
The weights for each of these characteristics can be set by the user, reflecting the aspect of tension  deemed most important. The integration of tension in the objective function of the optimization is discussed in detail in Section~\ref{sec:opt}. The next section focuses on the the pattern detection algorithm implemented in MorpheuS to improve long-term coherence.

\section{Detecting recurring patterns}
\label{sec:patterndetection}

Automatic recognition, description, classification and grouping of patterns are  important problems in many domains~\cite{jain2000statistical}. Applications include image segmentation~\cite{sclove1983application}, human action recognition~\cite{ji20133d}, face description~\cite{ahonen2006face}, DNA sequence analysis~\cite{gingeras1998simultaneous}, speech recognition~\cite{itakura1975minimum}, music genre recognition~\cite{tzanetakis2002musical}, and affective computing~\cite{picard1997affective}. We focus on pattern analysis for polyphonic music. 

When listening to a musical piece, a listener is able to recognize structure through perceiving repetition and relationships between parts of the piece of music. In order for a generated musical piece to sound natural, such patterns should exist. MorpheuS uses recurring patterns such as themes and motives from a template piece to fix these structural elements in a new composition. The detected patterns consist of groups of notes that can recur transposed in different places throughout the piece. There has been research on pattern detection techniques for music audio~\cite{dannenberg2003pattern, aucouturier2007bag}, but our focus is on symbolic music (i.e. MIDI). 

MorpheuS uses two state-of-the-art greedy compression-based algorithms for MIDI, COSIATEC and SIATECCompress~\cite{meredith2013cosiatec}, both based on Meredith's ``Structure Induction Algorithm'' (SIA) and SIATEC. SIA finds all the maximal translatable patterns (MTP) in a point-set and SIATEC discovers all translational equivalence classes (TECs) of MTPs in a point-set~\cite{meredith2002method}. The performance of both algorithms is benchmarked on a compression task in~\cite{meredith2013cosiatec}. The specific application of finding patterns for music generation requires special consideration when applying these algorithms. A discussion of the effect of parameter choices on the chosen pattern detection algorithm can be found in Section~\ref{sec:patternresults}. MorpheuS offers the user a choice of which algorithm to use as each has its own strengths and weaknesses.

\begin{figure}[ht!]
\centering
\begin{subfigure}[b]{1.0\columnwidth}
 
\includegraphics[width=\columnwidth]{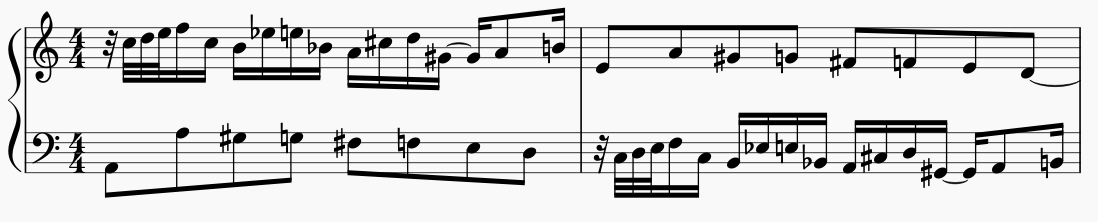}
\caption{First two bars of Bach's 20th prelude (Book II). }
\end{subfigure}

\vspace{0.1in}

\begin{subfigure}[b]{1.0\columnwidth}
 \includegraphics[width=\columnwidth]{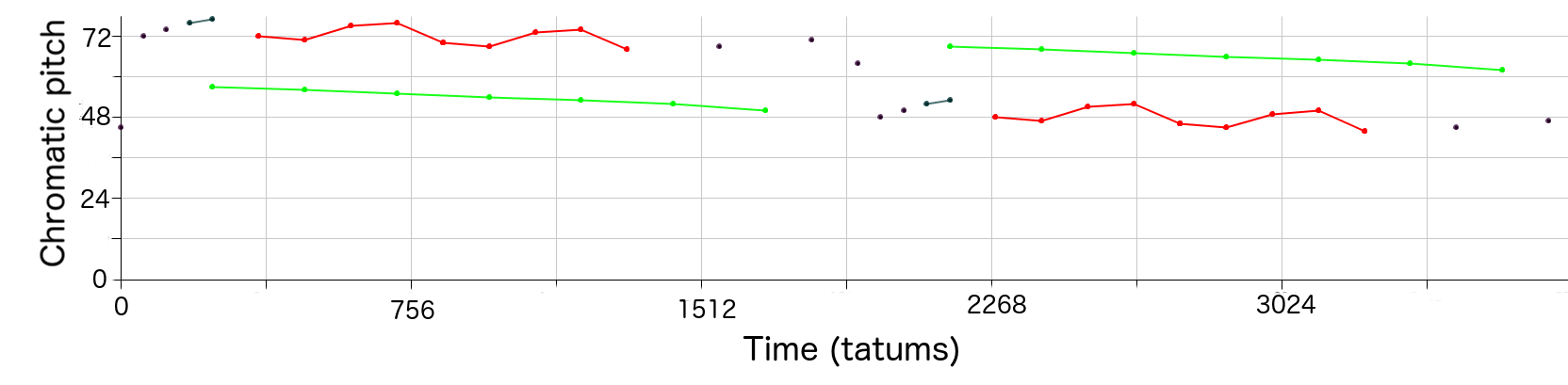}
 \caption{Patterns detected with COSIATEC~\cite{meredith2013cosiatec}.}
\end{subfigure}
\caption{COSIATEC applied to a short musical excerpt. Each TEC is represented with a different color~\cite{herremans2016morpheus}. }
\label{fig:patterns}
\end{figure}

When applied to polyphonic MIDI files, the compression algorithms use a point-set representation of the score, which positions each note in a two-dimensional pitch/time space. They then compute a compressed encoding, which takes the form of a set of TECs of maximal-length patterns. An example output of COSIATEC in pitch/time space is shown in Figure~\ref{fig:patterns}, whereby the time is expressed in tatums, i.e., ``the regular time division that mostly coincides with all note onsets''~\citep{bilmes1993timing}. Two longer patterns (displayed in red and green) are detected in the figure. A pattern (or a repetition of a pattern) is shown as a connected sequence of points, and its TEC consists of a musical transposition of the original pattern (one translation unit is a semitone). The two main patterns in the fragment recur, transposed, in the other hand. The red pattern, for instance, starts on the fifth note of the right hand (C), it recurs in the second bar (left hand) at the fifth note, transposed two octaves down. The encoded representation of the red (wavy) pattern in the figure is as follows:

\footnotesize
\begin{verbatim}
T(P(p(360,72),p(480,71),p(600,75),p(720,76),p(840,70)),
V(v(0,0),v(480,-2),v(1920,-24),v(2400,-26)))
\end{verbatim}
\normalsize
whereby the set of pairs \verb|P()| represents a maximal-length pattern, consisting of individual points \verb|p()| in pitch/time space. The set \verb|V()| contains the translation vectors \verb|v()|, which when applied to \verb|P()| form a translational equivalent pattern. The combination of the pattern and its translation vectors form \verb|T()|, a translational equivalence class of maximal-length patterns (MTP TEC). 

The first algorithm implemented in Morpheus, SIATECcompress, runs SIATEC once to get a list of MTP TECs and then selects a subset of TECs that covers the input dataset~\cite{meredith2015music}. The second algorithm, COSIATEC, on the other hand, iteratively uses SIATEC to find the best TEC, then removes this from the input dataset and repeats the process~\cite{meredith2013cosiatec}. Both algorithms result in a set of TECs with high compression ratios that cover a point-set. 
The encodings generated by COSIATEC are generally more compressed, meaning that the size of the file listing all TECs is smaller. SIATECCompress produces patterns that may intersect, which may be more relevant in the context of music analysis, as a note may belong to more than one musically meaningful pattern.
SIATECCompress performed best in the 2013 and 2014 MIREX competition on
``Discovery of repeated themes and sections'', and COSIATEC outperformed SIATECCompress on a Dutch folk song classification task with an accuracy rate of 84\% \cite{meredith2015music}.

The next section describes polyphonic music generation as an optimization problem which imposes the way the patterns repeat in the generated piece using hard constraints. This allows us to constrain the form and repetition structures of the newly generated piece. 






\section{Optimization problem}
\label{sec:opt}

In this research, generating music is modeled as an optimization problem. The main advantage of this is that it allows us to constrain global structure, consisting of repeated patterns, and to optimize the music to fit a tension profile. In this section, the resulting combinatorial optimization problem is formally defined.

\subsection{Variables}

The algorithm starts with a template piece whose rhythm and dynamics are treated as constants in the generated piece. The aim of MorpheuS is then to find a new set of pitches, $x$, for each note of the template piece, that minimizes the objective function and satisfies the repeated pattern constraints. 

\subsection{Objective function}

The objective of the optimization problem is to find a solution $x$ that matches a given tension profile as closely as possible. This tension profile can either be calculated from the template piece $t$ or could be manually input by the user. It comprises of three parts: one for each of the three tension measures $i \in \{0,1,2\}$ from Section~\ref{sec:tension}, represented as a vector $T_i(x)$ with length $l_i$. 
Since we want to match the tension profile of the solution $x$ to that of the template $t$, we calculate the Euclidean distance between these two tension profiles: 


\begin{equation} 
\label{eq:distance}
D_i(x) = \sum_{j = 1}^{l_i}\sqrt{(T_{ij}(x) - T_{ij}(t))^2}.
\end{equation}

The sum of the distances between each of the tension measures forms the objective function $D(x)$, which we aim to minimize. 

\begin{equation} 
\label{eq:distance2}
 D(x) = \sum_{i = 0}^{2} a_i \times D_i(x),
\end{equation}

\noindent
where $a_i$ is the weight for tension measure $i$. The weights offer the user a way to specify the relative importance of certain tension measures. In this paper, the weights are all set to 1. 

\subsection{Soft constraints} 

In addition to the hard constraints to be described in the next section, the user can elect to fix certain pitches in the solution. In order to do this, an additional term was added to the objective function $D(x)$ which imposes an arbitrary high penalty if $pitch(n_j)$ of note $j$ is not set to the required pitch ($setpitch(n_j)$): 

\begin{equation}
D'(x) = D(x) + b \times \sum_{j = 0}^{j}  C(n_j),
\end{equation}
whereby
\begin{equation*}
C(n_j)= \begin{cases}
    0, & \text{if $pitch(n_j) = setpitch(n_j)$}\\
    1, & \text{otherwise}
  \end{cases}
\end{equation*}
and $b$ is an arbitrary large number.

\subsection{Hard constraints} 

A number of the variables (pitches) of the solution $x$ are hard constrained to enforce the patterns detected in the template piece (as described in Section~\ref{sec:patterndetection}). This constraint ensures the recurrence of themes and motives in the output musical piece. The data structure used to store the solution is such that only the pitches of the original occurrence of the patterns \verb|p()| need to be decided. All other occurrences of a pattern are automatically set based on the pitches for the original pattern and the set of translational equivalence vectors \verb|V()| of the pattern. This setup speeds up the algorithm as it drastically reduces the size of the variables in the set $x$. 

In addition to the pattern constraints, an additional hard constraint is imposed on the pitch range for each track. This range is set based on the lowest and highest occurring pitch in the template piece for each track. Within this range, all possible pitches are allowed.



\section{Variable Neighborhood Search}


In this section, we describe the variable neighborhood search (VNS) algorithm used to solve the optimization problem defined above. Much of the research on the development of metaheuristics stems from more traditional fields such as vehicle routing and scheduling. In this research, we chose to implement a VNS algorithm because it has been shown to outperform several other heuristics (including genetic algorithms) on a range of  problems~\citep{hansen2001variable}. Since its inception in the late 90s, it has been successfully applied to problems in combinatorial optimization including project scheduling \citep{fleszar2004solving}, finding extremal graphs \citep{caporossi2000variable}, vehicle routing \citep{braysy2003reactive}, graph coloring \citep{avanthay2003variable}, and piano fingering~\cite{balliauw15}. 

A VNS algorithm has previously been developed for generating counterpoint music ~\cite{herremans2012composing,herremans2013composing}. This algorithm has proven to be efficient and outperformed a genetic algorithm implemented on the same problem. The inner workings of the algorithm have been modified to work with complex polyphonic piano music, and the constraints and objective function described in Section~\ref{sec:opt} have been integrated into the algorithm. 

 
\subsection{Local search components}

The core of a VNS algorithm is a local search strategy. Local search typically starts from an initial solution $x$, and iteratively makes a small change (i.e. a move) in order to find a better solution. We refer to the set of solutions $x'$ that can be reached by applying one type of move to a solution as the \emph{neighborhood}. In this case this means that the neighborhood will consist of all solutions that can be reached by applying one type of move to any of the time slices of the piece. A \emph{first descent strategy} was implemented in MorpheuS, whereby the neighborhood is built for one note/time slice at a time. As soon as a (feasible) solution is found that has a better value for the objective function $D(x')$, this solution is accepted as the new current solution $x$. 

An additional strategy for accelerating the search applies the moves chronologically from the start to the end of the piece. When a move is successful, this change will affect the tension profile only in its immediate vicinity. Therefore, the algorithm will backtrack only 4 time slices then resume the search.


\vspace{-.4cm}

\begin{figure}[h] \centering
\includegraphics[width=0.42\textwidth]{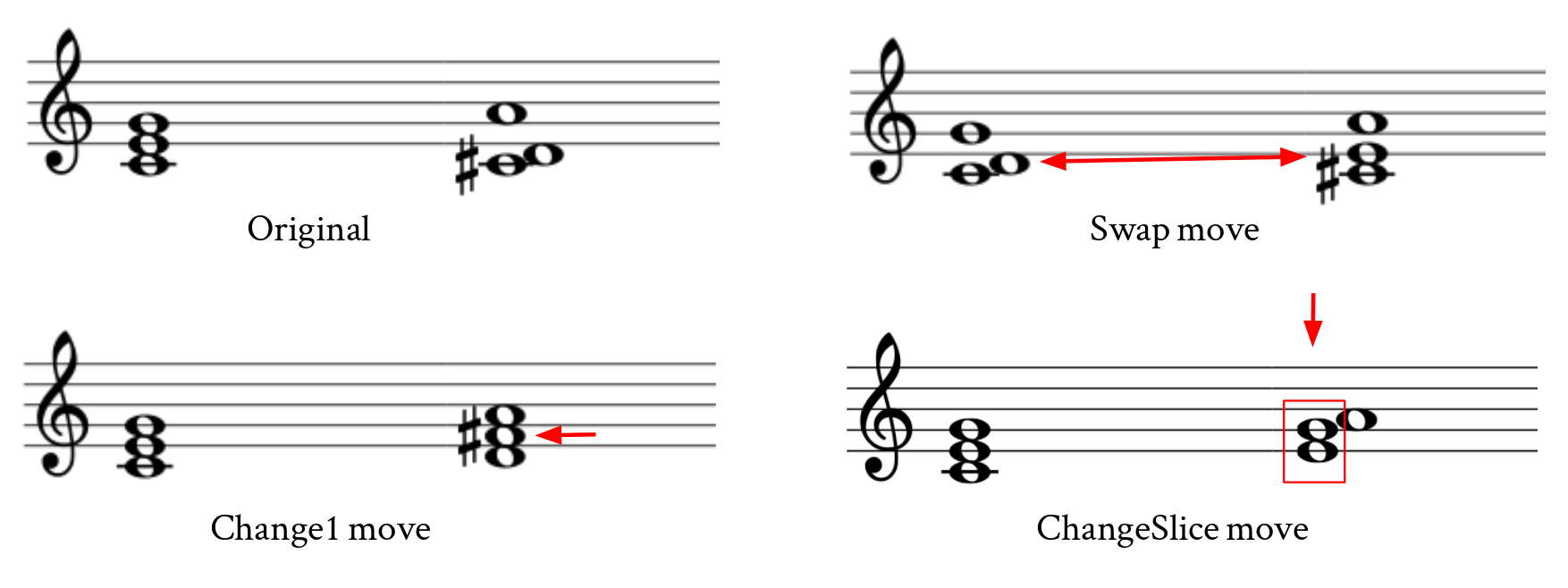}
\caption{An example of a potential move using each of the three different types of moves~\cite{herremans2016morpheus}. }
\label{fig:moves}
\end{figure}
\vspace{-.1cm}

Three types of moves are implemented in MorpheuS based on~\cite{herremans2012composing}. An example of each type of move is displayed in Figure~\ref{fig:moves} using a very short fragment. The \nbh{change1} move changes the pitch of one note to all of the other possible pitches in the allowed range to form the neighborhood. The \nbh{swap} move consists of all musical pieces that can be created by swapping the pitch of any two notes of the current piece. 
Finally, \nbh{changeSlice} changes the pitches of two randomly chosen notes in a vertical time slice to all of the other allowed pitches in the range. The respective size of each neighborhood generated by these three types of moves is displayed in Table~\ref{tab:size}. In order to speed up the algorithm, a first descent strategy is implemented, in which the neighborhood is built one move at a time. Whereas a steepest descent strategy would generate the full neighborhood, the first descent strategy accepts a new solution as soon as it improves the value of the current solution (see previous subsection).

\begin{table}[h]
\centering
\caption{Size of the neighborhood generated by each move type for a piece consisting of $n$ chords, each containing $m$ notes, and with a pitch range of $p$.}
\label{tab:size}
\begin{tabular}{lc}
\toprule
Move type & Neighborhood size \\
\midrule
\nbh{change1} & $m \times n \times p$\\
\nbh{swap} & $((n \times m)-1)!$ \\
\nbh{changeSlice} & $p^2$ \\
\bottomrule
\end{tabular}
\end{table}


\subsection{Outline of the VNS}

\begin{figure}\hspace{-1cm}
\includegraphics[width = .6\textwidth]{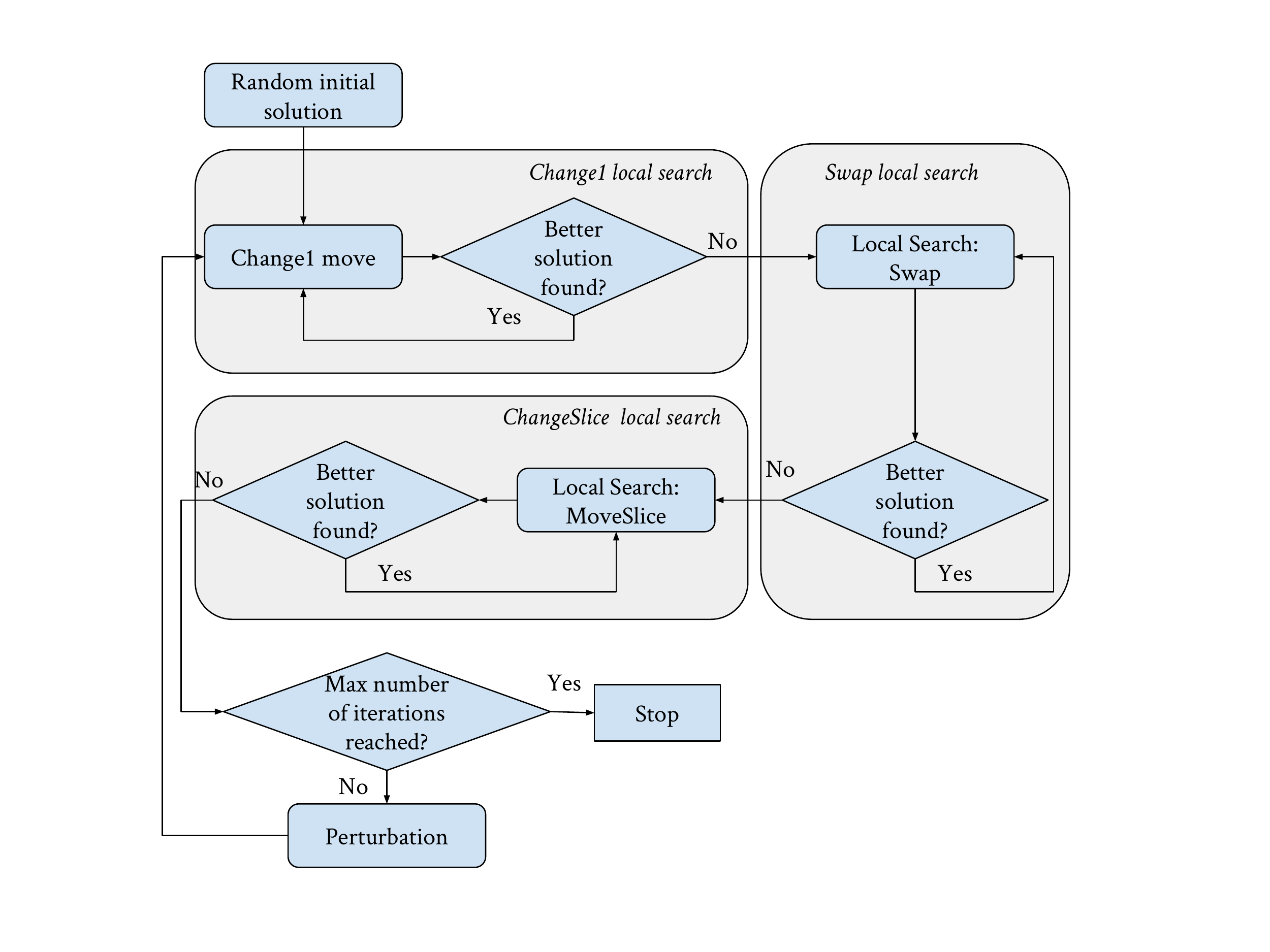}
\caption{Flow chart of the variable neighborhood search algorithm~\cite{herremans2016morpheus}.}
\label{fig:vns}
\end{figure}

A system diagram of the full algorithm implemented in MorpheuS is shown in Figure~\ref{fig:vns}. The VNS starts from a random (feasible) solution, which is built by assigning pitches from the range in a uniformly random manner. This initial solution is set as the current solution $x$.

The algorithm then performs local search using the \nbh{change1} neighborhood. When no improving feasible solution can be found, the VNS will then switch the local search to a different neighborhood type (e.g. \nbh{changeSlice}), which then allows the search to continue~\citep{mladenovi1997variable}. This process is repeated until no better solution can be found in any of the neighborhoods, in which case the algorithm is said to have arrived at a local optimum. 
The VNS algorithm implements a perturbation strategy to escape this local optimum and then continues the search for the global optimum~\citep{hansen2003variable}. A perturbation re-assigns a significant proportion of the pitches each to a uniform random (feasible) pitch. Based on the research of~\citet{herremans2013composing}, the number of perturbed pitches was set to 12\%. In an iterative local search strategy, the algorithm restarts from a totally random solution when a local optimum is reached. The perturbation strategy implemented in the VNS, however, leads to far better results~\citep{lourenco2003iterated}. The search process continues until the stopping criterion (i.e. a maximum allowed number of iterations) is reached. 
The order in which the different types of moves are applied is based on the increasing computational complexity of calculating the full neighorhood. 
In the next section, we evaluate the implemented algorithm and its musical results.


\section{Results}

The MorpheuS system is evaluated on three levels. The first examines the effects of pattern detection on the musical outcome. Next, we consider the efficiency of the optimization algorithm. Last but not least, the generated musical output is evaluated and compared to the original template piece.

\subsection{Effect of pattern detection algorithm}
\label{sec:patternresults}

\begin{figure*}[hbt!]	\centering	
\includegraphics[width=.8\textwidth, trim={0cm 8cm 0cm 2cm}, clip]{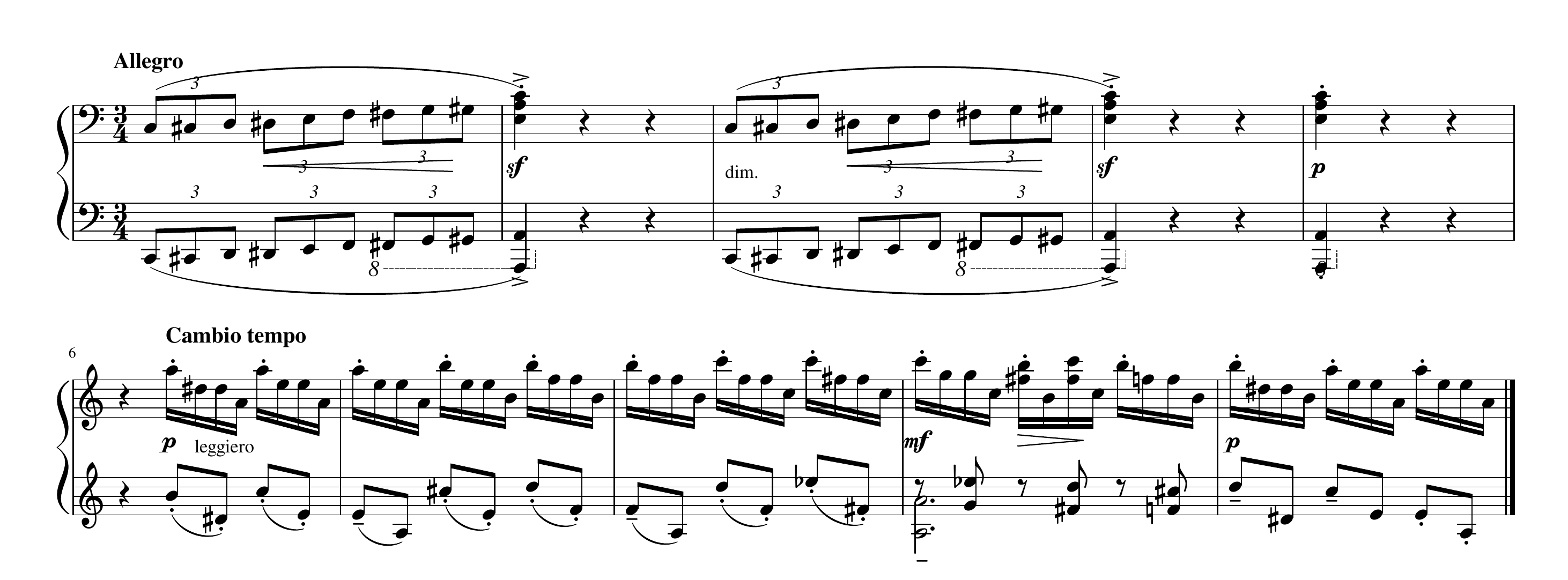}
\caption{Original excerpt from Rachmaninov's ``\'Etude Tableau Op. 39, No. 6''}
\label{fig:rach3}
\end{figure*}

The selected pattern detection algorithm (COSIATEC versus SIATECCompress) exerts a big influence on the resulting pieces. Short but frequent patterns can overly constrain the generation process, thus forcing it to converge quickly to the original piece. Infrequently repeated patterns, even though they may be long, easily lose sight of the goal of constraining long term structure. The user can specify which algorithm is implemented: COSIATEC, which uniquely captures each note in precisely one pattern, or SIATECCompress, which captures more relationships between different notes, resulting in overlapping patterns. Each of these algorithm in turn has additional settings, such as maximum and minimum pattern lengths. We have generated three different pattern sets based on an excerpt of Rachmaninov's ``\'Etude Tableau Op. 39, No. 6'', shown in Figure~\ref{fig:rach3}.

\begin{figure*}[h]
\begin{subfigure}[h]{\textwidth}
		\centering
		\includegraphics[height=2.6cm, width=.8\textwidth]{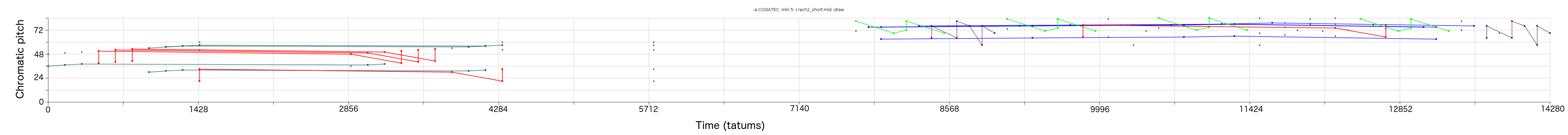}
		\caption{Pattern set A: COSIATEC, minimum pattern length 5, compression ratio: 1.65, Number of TECs: 6}
        \label{fig:pa1}	
	\end{subfigure}
    
        \begin{subfigure}[h]{\textwidth}
		\centering
 		\includegraphics[height=2.6cm, width=.8\textwidth]{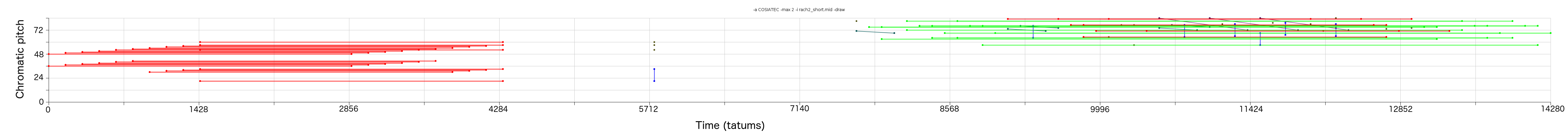}
		\caption{Pattern set B: COSIATEC, maximum pattern length 2, compression ratio: 1.67, number of TECs: 6}
        \label{fig:pa3}
	\end{subfigure}
    
\begin{subfigure}[h]{\textwidth}
		\centering
 		\includegraphics[height=2.6cm, width=.8\textwidth]{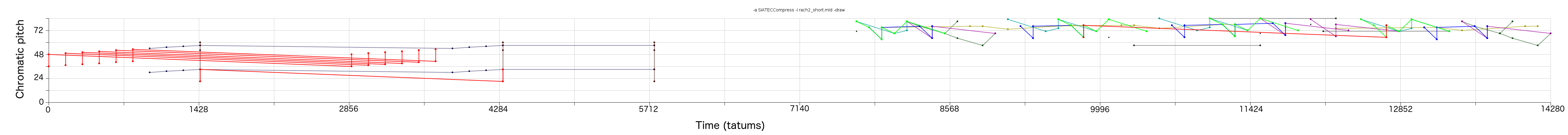}
		\caption{Pattern set C: SIATECCompress (no restrictions on the pattern length), compression ratio: 1.58, number of TECs: 11. Each TEC is represented with a different color.}
        \label{fig:pa2}
	\end{subfigure}
    
\caption{Different patterns detected in  Rachmaninov's ``\'Etude Tableau Op. 39, No. 6''}
\label{fig:patternsRach3}
\end{figure*}

Based on this excerpt, we have calculated three sets of repeated patterns, as displayed in Figure~\ref{fig:patternsRach3}. Their main properties, namely, compression ratio, number of notes in patterns, \verb|P()|, and the size of the TECs, are shown in Table~\ref{tab:patterns}. Each of these three pattern sets was then used as a structural template during music generation in MorpheuS. The resulting music pieces generated, based on a short run of 10 iters (less then 1 minute generation time on a Dell XPS13 laptop with i7core and 8GB RAM) are displayed in the figure in Appendix~A. An example of each detected pattern, together with the set of translational equivalent vectors is shown on each score in green and orange respectively.  

The first pattern set (A), shown in Figure~\ref{fig:patternsRach3}(a), was detected using COSIATEC with a minimum pattern length of 5. This resulted in 6 TECs with a compression ratio of 1.65, and 69 unique notes that needed to be optimized by MorpheuS (see Table~\ref{tab:patterns}). The resulting piece, created by iterating through the VNS 10 times is displayed in (a) of Appendix~A. The music retains some of the contours of the original piece, but also contains a great deal of original musical content. 

When constraining COSIATEC to detect only very short patterns of maximum length 2, we obtain a set of TECs (B), shown in Figure~\ref{fig:patternsRach3}(b). This yields a very similar compression ratio, yet the piece generated based on this template pattern is very different. In this case, the original piece is almost replicated exactly due to the many constraints posed by the set of TECs. The prevalence of such short patterns typically severely limit the originality of the music generated.  Here, MorpheuS only has 11 notes to optimize; the others were derived from the translation vectors of the pattern vector.

Pattern set C, shown in Figure~\ref{fig:patternsRach3}(c), shows the results of running SIATECCompress without any constraints on pattern length. Although the compression ratio of COSIATEC is often higher than that of SIATECCompress~\cite{meredith2015music}, musically speaking, the latter may have further benefits. In SIATECCompress, a note can be contained in multiple sets of TECs, which allows the algorithm to find a different and larger set of TECs than COSIATEC. This may result in the algorithm capturing more meaningful musical relationships. The resulting music generated using pattern set C as a template offers a mix between the highly constrained nature of pattern set B and the freedom of pattern set A. An example of this can be found in the ascending pattern in bars 1 and 3. In pattern set C, both bars have an ascending line, yet the starting note is different. Pattern set B generates a more constrained output, whereby both bars are identical. With pattern set A, we see a much freer interpretation, whereby the two bars bear minimal resemblance to each other. Although each of the three example patterns offer a way to constrain long-term structure in generated music, the degree to which they constrain pitch patterns has significant effect on the musical outcome. 

\begin{table}[h]
\centering
\caption{Pattern sets generated for Rachmaninov's ``\'Etude Tableau Op. 39, No. 6''}
\label{tab:patterns}
\begin{tabular}{lllll}
\toprule
       & Algorithm      & CR & UP & TECs \\
       \midrule
Pattern set A & COSIATEC       & 1.65              & 69                                             & 6              \\
Pattern set B & COSIATEC       & 1.67              & 11                                             & 6              \\
Pattern set C & SIATECCompress & 1.58              & 34                                             & 11   \\     
\bottomrule
\end{tabular}

\vspace{.2cm}
\scriptsize CR: compression ratio, UP: number of pitches to be set by MorpheuS
\end{table}

\subsection{Evolution of solution quality}

A formal comparison with other existing systems was not possible due to the fact that MorpheuS is the first algorithm that implements a tension based objective function with pattern constraints. The use of VNS for generating counterpoint, on which MorpheuS is based, has been tested extensively in~\cite{herremans2012composing} and shown to outperform a genetic algorithm on the same task. We can thus assume that, in the more constrained (due to the patterns imposed) musical task represented here, the algorithm will be at least as effective. 

\begin{figure}[hbt!]	\centering
\begin{subfigure}[h]{.45\textwidth}
\includegraphics[width=1\columnwidth]{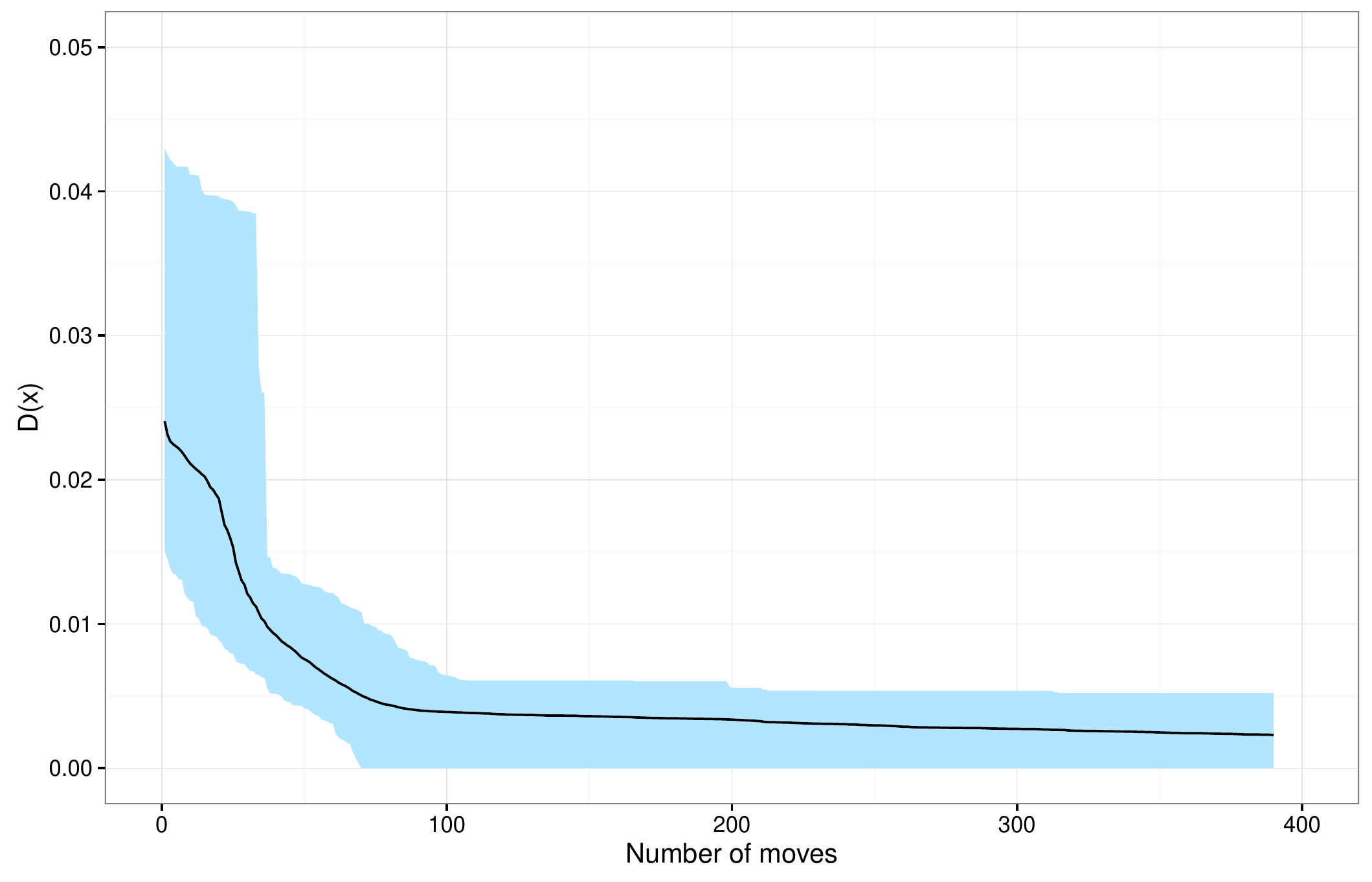}
\caption{For Rachmaninov's ``\'Etude Tableau Op. 39, No. 6'' (with Pattern set C)}
\label{fig:rachvns100}
\end{subfigure}
\begin{subfigure}[h]{.45\textwidth}
\includegraphics[width=1\columnwidth]{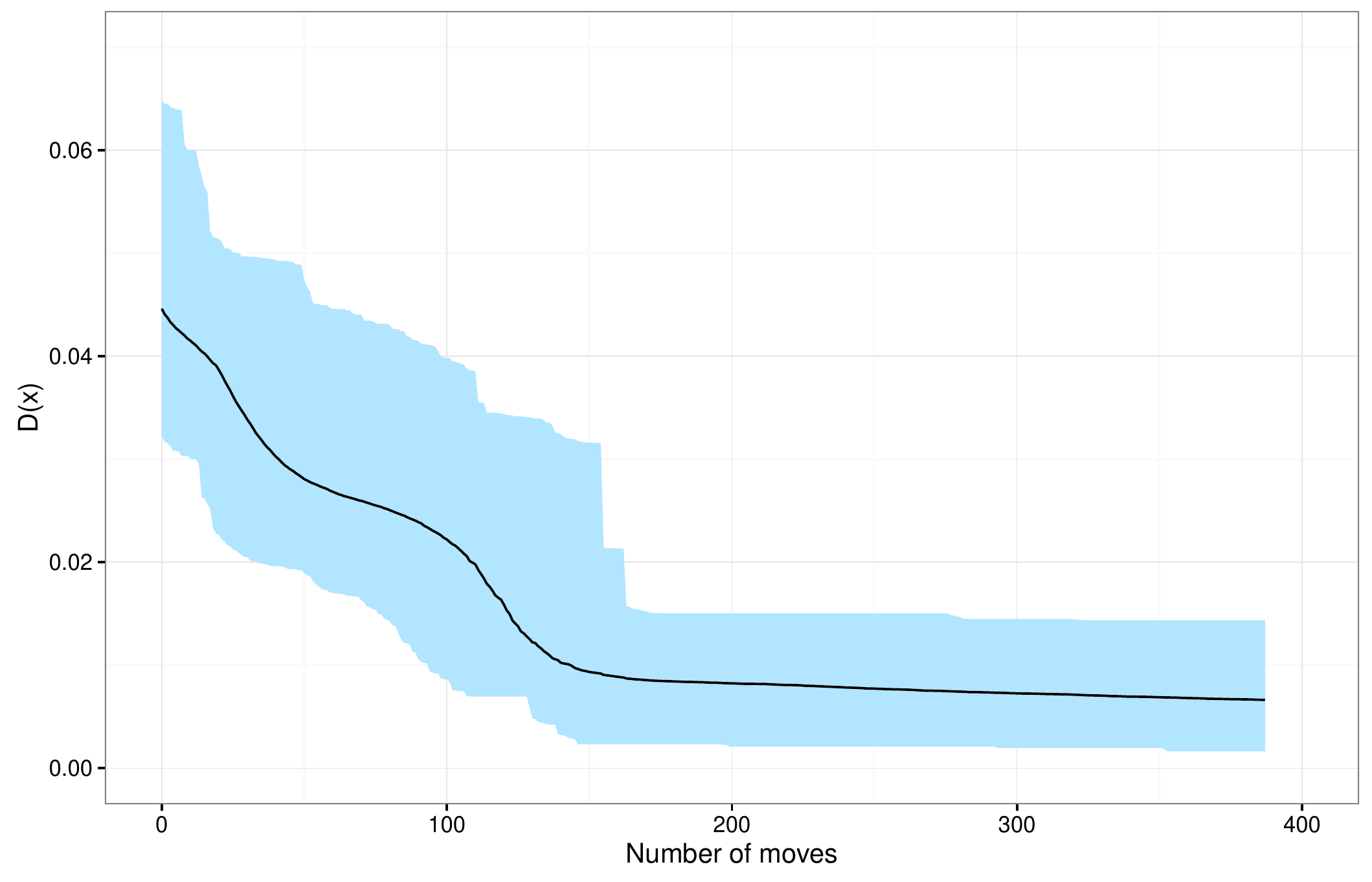}
\caption{Kabalevsky's ``Clowns''.}
\label{fig:clownsvns100}
\end{subfigure}
\caption{Evolution of the objective function over time, for 100 runs of the VNS. The plotted line shows the mean best solution found by the VNS for 100 runs. The ribbons show the maximum and minimum objective function values for the best solution found over the 100 runs at each move.}
\label{fig:100runs}
\vspace{-.4cm}
\end{figure}

In order to verify the effectiveness of the algorithm it was run 100 time on Kabalevsky's  ``Clowns'' (from \textit{24 Pieces for Children}, Op. 39 No. 20) with SIATECCompress patterns, and the Rachmaninov ``\'Etude Tableau Op. 39, no. 6'' shown above (using pattern set C). Figure~\ref{fig:100runs} shows the range and average of the best objective function value found over time for 100 runs of the VNS for both pieces. The experiment was performed on a Dell XPS Ultrabook with i7Core and 8GB RAM. The average running time of the VNS was 136 seconds for ``Clowns'' and 526 seconds for the Rachmaninov piece. The size of the solution was 34 and 84 notes, respectively.

Figures~\ref{fig:rachvns100} and~\ref{fig:clownsvns100} clearly show a steep improvement during the initial seconds of the algorithm's run for both pieces. This pattern can be observed for each of the 100 runs, as the maximum values for the best solution found (i.e. worst run of the algorithm), goes down quickly. Even after 2 minutes
, the algorithm manages to find small improvements to the current solution.

\begin{figure}[hbt!] \centering
\includegraphics[width=0.8\columnwidth, trim=0.1cm 0cm 0cm 0cm, clip]{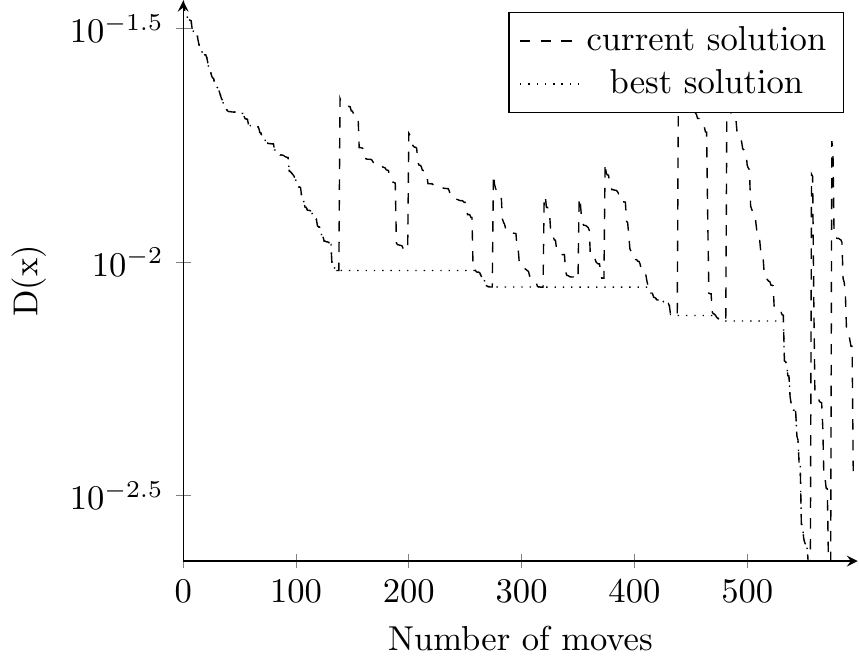}
\caption{Evolution of the objective function over time for one run of the VNS, for Kabalevsky's ``Clowns''.}
\label{fig:clownsvns}
\end{figure}

In Figure~\ref{fig:clownsvns}, we isolate one particular run of the algorithm on ``Clowns''. A clear descending trend can be observed when looking at the best solution found over time. The peaks in the graph indicate points of perturbation. Whenever the search gets trapped in a local optimum, the current solution is perturbed, leading to a temporarily worse solution. Note that even after 500 moves, the perturbation step manages to escape from a local optimum to find a better solution, thus confirming that the perturbation strategy is successful.

\begin{figure*}[hbt!] \centering
     \begin{subfigure}[h]{.49\textwidth}
		\includegraphics[width=1\textwidth, clip, trim={7cm 0 5cm 0}]{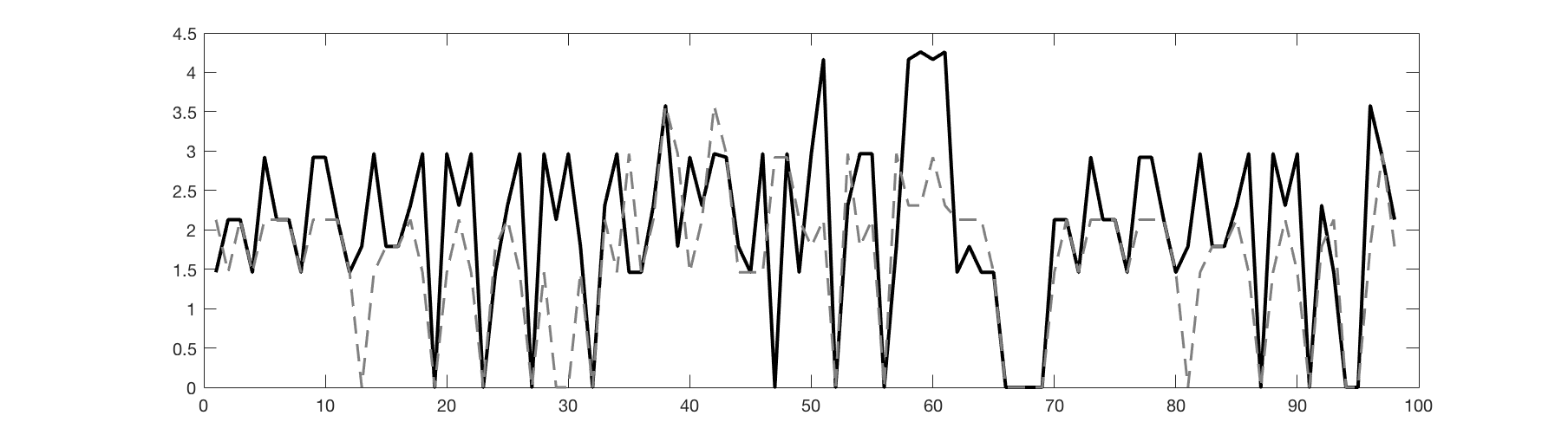}
		\caption{Cloud diameter: random (solid) \& original piece (dashed)}
        \label{fig:key_random}	
	\end{subfigure}
        \begin{subfigure}[h]{.49\textwidth}
		\includegraphics[width=1\textwidth, clip, trim={7cm 0 5cm 0}]{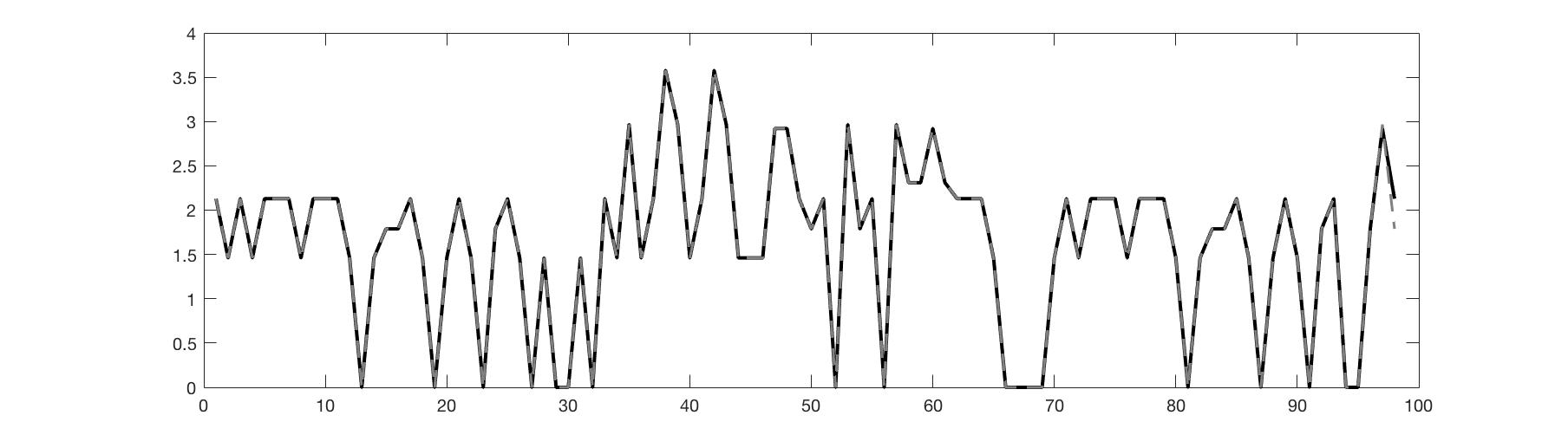}
		\caption{Cloud diameter: optimized (solid) \& original piece (dashed)}
        \label{fig:mom_random}	
	\end{subfigure}    

    \begin{subfigure}[h]{.49\textwidth}
		\includegraphics[width=1\textwidth, clip, trim={7cm 0 5cm 0}]{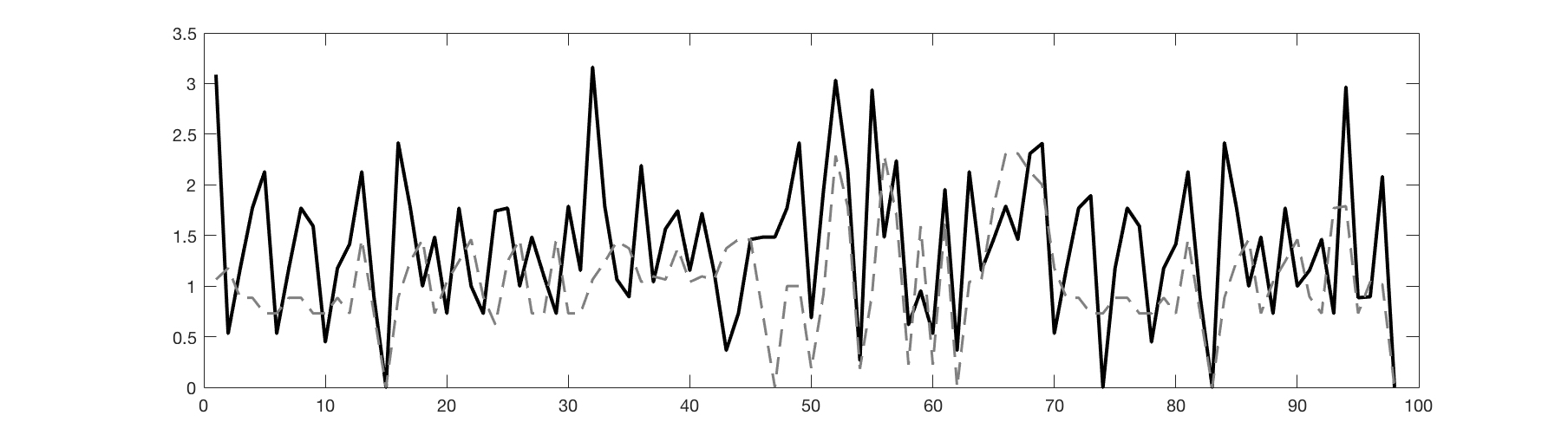}
		\caption{Cloud Momentum: random (solid) \& original piece (dashed)}
        \label{fig:key_start}	
	\end{subfigure}
        \begin{subfigure}[h]{.49\textwidth}
		\includegraphics[width=1\textwidth, clip, trim={7cm 0 5cm 0}]{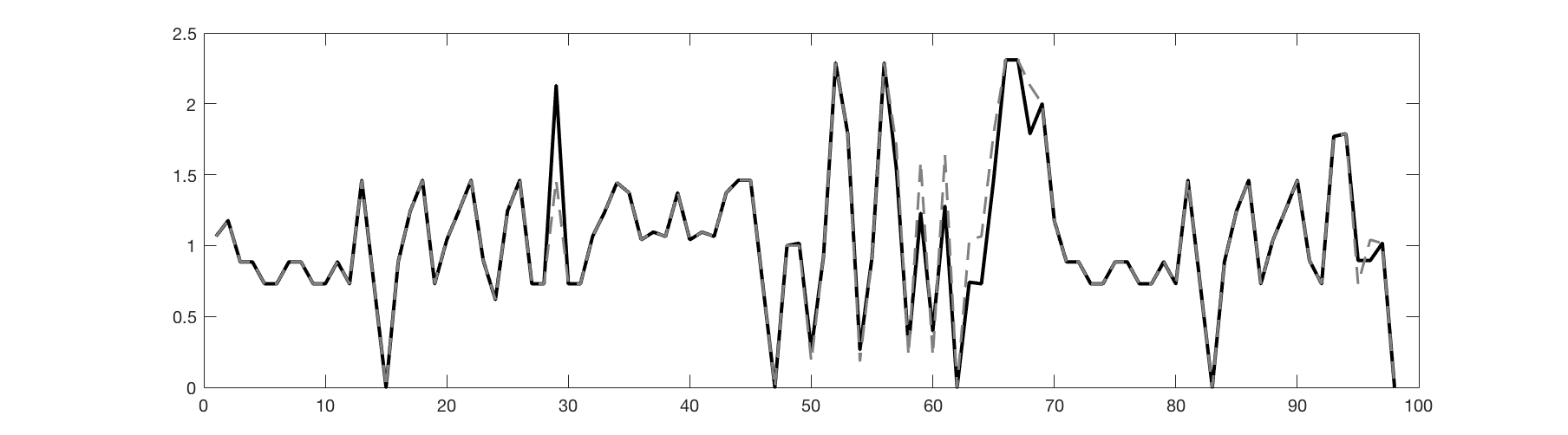}
		\caption{Cloud momentum: optimized (solid) \& original piece (dashed)}
        \label{fig:mom_start}	
	\end{subfigure}
%
\begin{subfigure}[h]{.49\textwidth}
		\includegraphics[width=1\textwidth, clip, trim={7cm 0 5cm 0}]{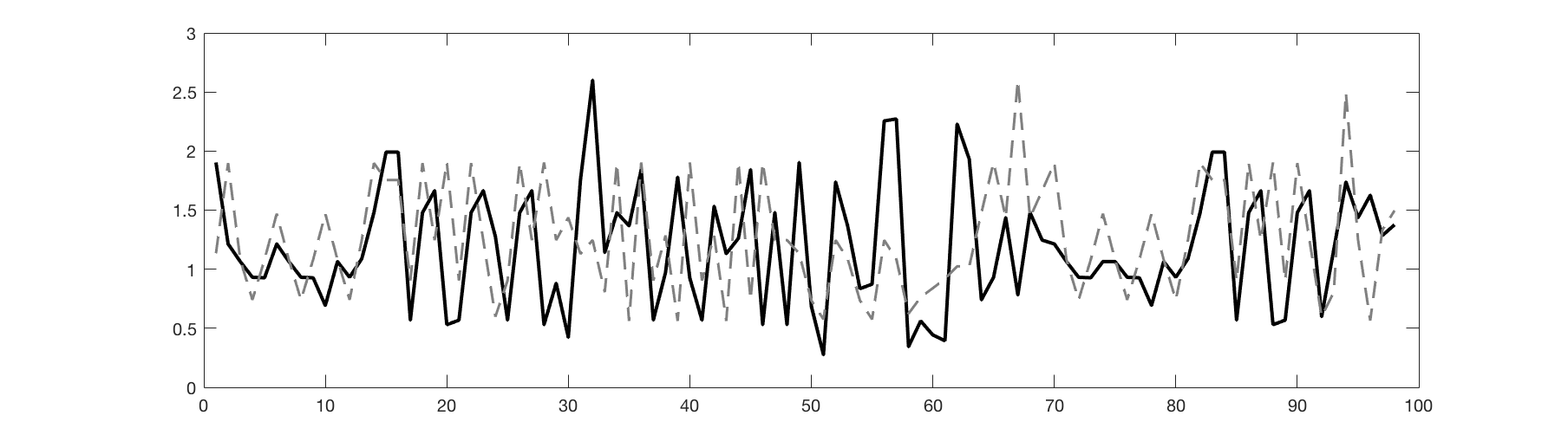}
		\caption{Tensile strain: random (solid) \& original piece (dashed)}
        \label{fig:key_end}	
	\end{subfigure}
\begin{subfigure}[h]{.49\textwidth}
		\includegraphics[width=1\textwidth, clip, trim={7cm 0 5cm 0}]{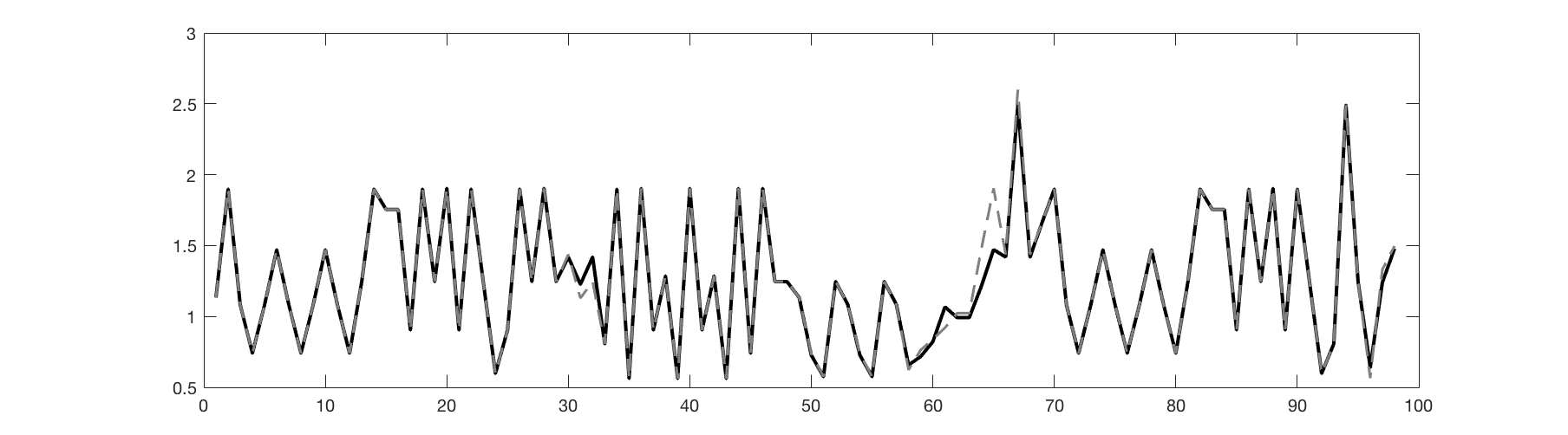}
		\caption{Tensile Strain: optimized (solid) \& original piece (dashed)}
        \label{fig:mom_end}	
	\end{subfigure}
    \caption{The three types of tension profiles before and after optimization of Kabalevsky's ``Clowns'', together with the template profile (dashed). The y-axes represent the value of each tension score (cloud diameter, tensile strain and cloud momentum), and the x-axes represent score time. 
    }
    \label{fig:tensionclowns}
    \vspace{-.4cm}
\end{figure*}

The three tension profiles: cloud diameter, tensile strain and cloud momentum, are shown in Figure~\ref{fig:tensionclowns}. 
The graphs show the original tension profile of the template piece (dashed line), and snapshots of the tension profiles of the output piece before optimization (random piece) and after optimization. It can be noticed that the tension profile of the original piece fluctuates between tension and relaxation, a dynamic that~\citet{lerdahl1987generative} discuss in their Generative Theory of Tonal Music. 

The graphs of the random piece, however, show a much more erratic tension profile distant from that of the original piece. Overall, the tension is also higher for the random pieces, which can in part be explained by the dissonance that we can expect in pieces with random pitches. A striking similarity can be seen between the tension profiles of the original piece (template) and the optimized piece, yet again confirming that the optimization algorithm indeed finds a solution that minimizes the objective function. This is confirmed by the correlation coefficients, which are high between the tension profiles of the optimized and template piece (0.9748, 0.9918, 0.9993), and lower between the random initial solution and the template piece (0.4103, 0.2053, and 0.6174). In the next section we will focus on the actual musical results.

\subsection{Musical outcome}




It must be noted that in the examples given in this paper, the goal of the optimization is to fit the original tension profile of the \emph{template} piece as closely as possible. This explains a tendency to revert to the original piece, but offers us a way to verify that the optimization algorithm performs well. A future version of MorpheuS may include a `similarity-penalty' in the objective function, to enforce originality in the generated pieces.  Currently, the user is free to design their own tension profile to create an original piece, or use the tension profile of a template piece. 

Appendix~B shows the musical output of the optimization process described in the previous section together with the initial (random) starting piece, based on the template piece ``Clowns'' by Kabalevsky. A significant improvement in musical quality can be noticed between the random and the optimized piece. One of the most evident aspects is that the latter (optimized) piece is much more tonal. We can also see some long-term recurrent structures; for example, the theme from bars 1-5 returns in bars 18-22. 

The reader is invited to listen to this and other output generated by MorpheuS online\footnote{\url{http://dorienherremans.com/morpheus}}
While these first tests are promising, there is still room for improvement. One interesting improvement is to add constraints on playability. While the current pitch range constraint ensures that assigned notes occur within the range of the template piece, the output can be far from idiomatic for the instrument and variables such as the unexpectedness of the note sequences can make them difficult to play. Improved playability could be achieved by a statistical machine learning approach (e.g. Markov model or recurrent neural network) to integrate transition probabilities in the current objective function. 


Following live performances of MorpheuS' pieces, we have received a range of comments from expert musicians, reflecting interesting perspectives on MorpheuS' compositions to inform future developments of the system. Upon hearing MorpheuS' version of a Haydn sonata movement, an academic composer remarked that the system shows some competence with the repetition of material, however it does not develop the material; furthermore, it does not know (like Stravinsky) to use the `right' kind of `wrong' notes; MorpheuS' use (or misuse) of cadences, cadential figures were inserted in odd places, were the most obvious anomalies distinct from human composition; also, the evolution of harmony with respect to phrase structure, does not work as the tension levels do not relate to the phrase structure.  

Several expert listeners remarked on the humor apparent in MorpheuS' pieces, particularly the ways in which it naively violates conventional expectations, often with surprising and sometimes adventurous but plausible results. The advantage of this naivet\'{e}, as one person puts it, is that the system ``has no fear'' and thus ``has courage to try things'', unencumbered by conditioning that constrains human composers to behave or make choices only in ways that are deemed acceptable. This was in the context of three morphed pieces by Bach and three morphed pieces by Kabalevsky. In a few of these pieces, there were awkward harmonic moments when returning to the beginning of a repeated section, leading one expert listener to comment that MorpheuS lacked the ability to make good transitions. However, the listener found it fascinating to hear the original Bach pieces (from ``A Little Notebook for Anna Magdalena'') through the lens of MorpheuS' compositions. In contrast, another expert listener found the Kabalevsky pieces ``more honest'' than the Bach ones, likely because the original Bach pieces were too recognizable in the morphed ones, yet lacked certain characteristics typically associated with the pieces.

\section{Conclusions}

MorpheuS consists of a novel framework that allows us to tackle one of the main challenges in the field of automatic music generation: long-term structure. An efficient variable neighborhood search algorithm was developed that enables the generation of complex polyphonic music, with recurring patterns, according to a tension profile. Two pattern detection techniques extract repeated and translated note patterns from a template piece. This structure is then used as scaffolding to constrain long-term structure in new pieces. 
The objective function of the optimization problem is based on a mathematical model for tonal tension. This allows for the generation of pieces with a predefined tension profile, which has potential applications to game or film music.  The pieces generated by MorpheuS have proved to be promising and have been tested in live performance scenarios at concerts internationally. 

In future research it would be interesting to explore the integration of machine learning methods, e.g. deep learning techniques or Markov models, in the objective function of the optimization algorithm. This way, a style could be learned from a large corpus of music, and in turn be reflected in the generated piece. We expect that this would also improve the playability of the pieces and reduce awkward transitions. A second expansion would be to allow for more flexible pattern detection, such as recognition of inverted patterns, augmentations, and diminutions and other variations. It would equally be interesting to evaluate if the generated music elicits the same emotion responses as expected given a tension profile, by measuring physiological responses or by recording listener judgments of tension as described in~\cite{kim2008emotion,agres2015creativity}. The tension model could also be expanded to capture other characteristics of tension such as timbre and cadence. 

With regard to interaction design, it would also be interesting to test a transformational approach in which the optimization starts from the original piece and searches for a new piece that matches a tension profile provided by the user, thus transforming the original piece. Finally, in the context of adaptive game music generation, the VNS algorithm could be modified to allow for real-time generation, much like the system \citet{antor13} implemented as the FuX 2.0 mobile music generation app.


\ifCLASSOPTIONcompsoc
  \section*{Acknowledgments}
\else
  \section*{Acknowledgment}
\fi  This project has received funding from the European Union’s Horizon 2020 research and innovation programme under grant agreement No 658914.
We are grateful to Prof. Dr. David Meredith for providing us with implementations of his COSIATEC and SIATECCompress algorithms.

Finally, we thank the expert listeners who provided anecdotal feedback following performances of MorpheuS' output. They were Dr. Uzial Awrat, Dr. Oded Ben-Tal, Dr. Paul Edlin, and Carla Townsend Sturm.





\bibliographystyle{IEEEtranN}
%



%

\begin{IEEEbiography}[{\includegraphics[width=1in,height=1.25in,clip,keepaspectratio]{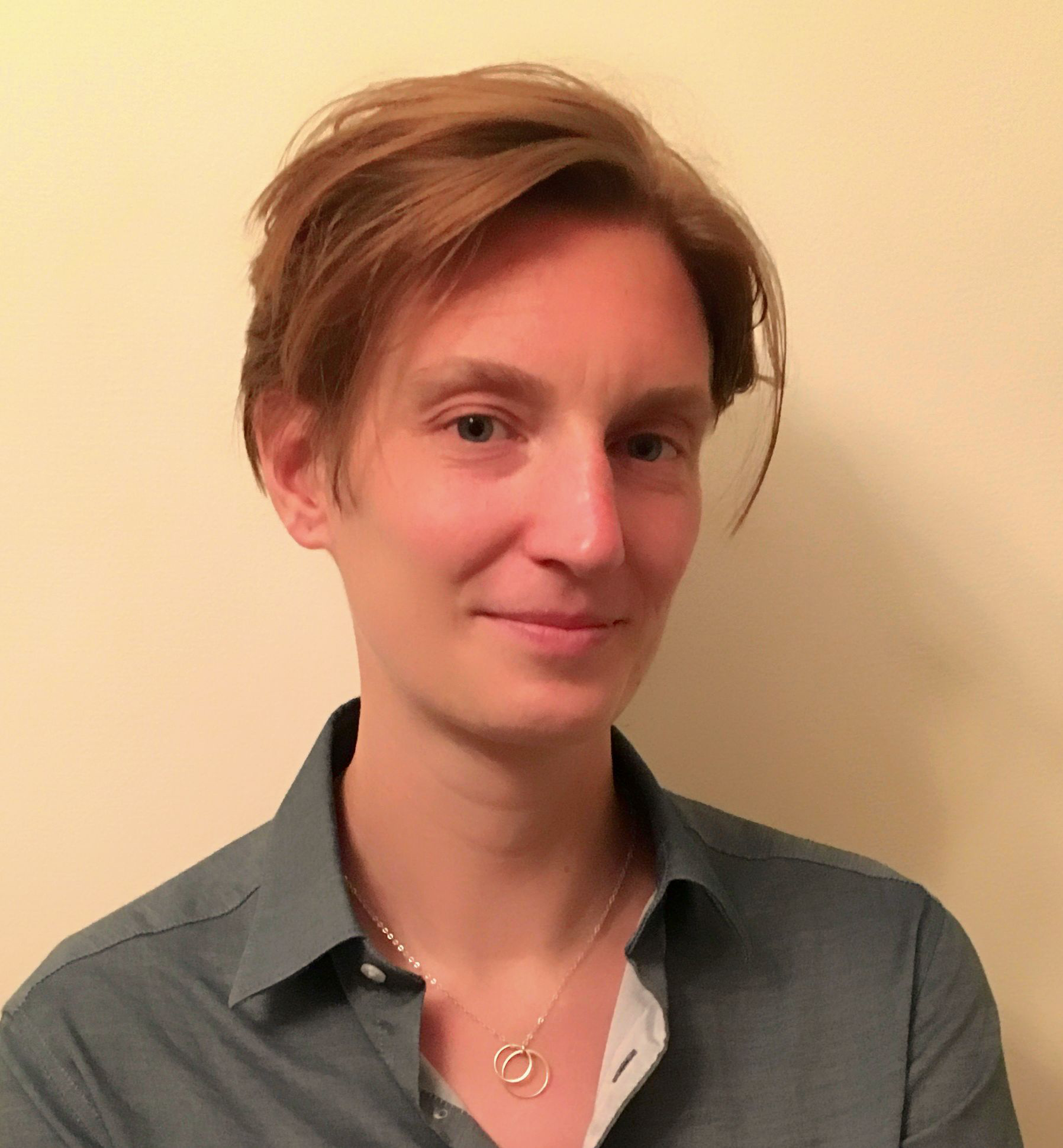}}]
{Dorien Herremans} is an Assistant Professor at the Information Systems Technology and Design Pillar at the Singapore University of Technology and 	Design, with a joint appointment at the Institute of High Performance Computing at the Agency for Science Technology and Research (A*STAR). In 2015, she was awarded the individual Marie Sk\l{}odowska-Curie Fellowship for Experienced Researchers, and worked at the Centre for Digital Music, Queen Mary University of London, on the project: ``MorpheuS: Hybrid Machine Learning – Optimization techniques To Generate Structured Music Through Morphing And Fusion''. 
Prof. Herremans received her PhD in Applied Economics from the University of Antwerp. Her PhD thesis was titled ``Compose$\equiv$Compute: Computer Generation and Classification of Music through Operations Research Methods''. After graduating as a commercial engineer in management information systems at the University of Antwerp in 2005, she worked as a Drupal consultant and was an IT lecturer at Les Roches University in Bluche, Switzerland. She also worked as a mandaatassistent at the University of Antwerp, in the domain of operations management, supply chain management and operations research, and was a visiting researcher at the Department of Computer Science and Artificial Intelligence at the University of the Basque Country, San Sebasti\'an. Prof. Herremans' research focuses the intersection of machine learning/optimization and digital music. She is a Senior Member of the IEEE and co-organizer of the First International Workshop on Deep Learning and Music as part of IJCNN. 
\end{IEEEbiography}


\begin{IEEEbiography}[{\includegraphics[width=1in,height=1.25in,clip,keepaspectratio]{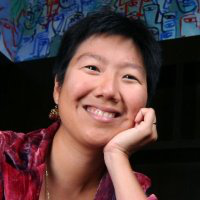}}]
{Elaine Chew} 
is Professor of Digital Media in the School of Electronic Engineering and Computer Science at Queen Mary University of London (QMUL) where she is affiliated with the Centre for Digital Music. Prior to joining QMUL in 2011, she was a tenured associate professor at the University of Southern California, where she was the inaugural holder of the Viterbi Early Career Chair. She was a recipient of the US Presidential Early Career Award in Science and Engineering and NSF CAREER Award, and the Edward, Frances, and Shirley B. Daniels Fellow at the Radcliffe Institute for Advanced Study. She is also an alum of the NAS Kavli Frontiers of Science and NAE Frontiers of Engineering Symposia. Her research centers on the mathematical and computational modeling of music structure, musical prosody, music cognition, and ensemble interaction. She is author of over 100 peer-reviewed chapters and articles, and author and editor of 8 books and journal special issues on music and computing. She has served as program and general chair of the International Conferences on Music Information Retrieval (2008) and of Mathematics and Computation in Music (2009, 2015), and was invited convenor of the Mathemusical Conversations international workshop in 2015. She was awarded PhD and SM degrees in operations research at the Massachusetts Institute of Technology, and a BAS in mathematical and computational sciences (hon) and music (distinction) at Stanford University.
\end{IEEEbiography}




\vfill



\footnotesize
\bibliography{paper}

\end{document}